\documentclass[10pt,journal,compsoc]{IEEEtran}

\usepackage{bm}

\usepackage{amsfonts} 
\usepackage{amsmath}
\usepackage{xcolor}
\usepackage{booktabs} 
\usepackage{multirow}
\usepackage{calc}
\usepackage{subcaption}
\usepackage[ruled,linesnumbered]{algorithm2e}
\usepackage[noend]{algpseudocode}
\usepackage[printonlyused]{acronym}
\newacro{DPM}{Dynamic Power Management}
\newacro{OFDM}{Orthogonal Frequency-Division Multiplexing}
\newacro{LTE}{Long Term Evolution}
\newacro{eNodeB}{evolved Base Station}
\newacro{PRB}{Physical Resource Block}
\newacro{UE}{User Equipment}
\newacro{RF}{Radio Frequency}
\newacro{AS}{Access Stratum}
\newacro{NAS}{Non Access Stratum}
\newacro{TTI}{Transmission Time Interval}
\newacro{PSM}{Power State Machine}
\newacro{PDCCH}{Physical Downlink Control Channel}
\newacro{PDSCH}{Physical Downlink Shared Channel}
\newacro{PUSCH}{Physical Uplink Shared Channel}
\newacro{FNR}{False Negative Rate}
\newacro{FPR}{False Positive Rate}
\newacro{BSR}{Buffer Status Report}
\newacro{PUCCH}{Physical Uplink Control Channel}
\newacro{PUSCH}{Physical Uplink Shared Channel}
\newacro{FLOP}{Floating Point Operation}
\newacro{DSP}{Digital Signal Processor}
\newacro{HTTP}{Hypertext Transfer Protocol}
\newacro{DLG}{Downlink Grant}
\newacro{ULG}{Uplink Grant}
\newacro{MCS}{Modulation and Coding Scheme}

\usepackage[capitalize]{cleveref}

\usepackage{tikz}
\usetikzlibrary{decorations.pathreplacing}
\usepackage{tikzscale}
\usepackage{pgfplots}
\usepackage[outline]{contour}

\definecolor{butter1}{rgb}{0.988,0.914,0.310}
\definecolor{chocolate1}{rgb}{0.914,0.725,0.431}
\definecolor{chameleon1}{rgb}{0.541,0.886,0.204}
\definecolor{skyblue1}{rgb}{0.447,0.624,0.812}
\definecolor{plum1}{rgb}{0.678,0.498,0.659}
\definecolor{scarletred1}{rgb}{0.937,0.161,0.161}
\definecolor{ao}{rgb}{.35, .447, .19}

\newcommand{\comm}[1]{}
\newcommand{\true}{\mathbf{t}}
\newcommand{\false}{\mathbf{f}}
\newcommand{\unknown}{\mathbf{u}}

\renewcommand{\vec}[1]{\mathbf{#1}}

\DeclareMathOperator{\argmax}{argmax}

\ifCLASSOPTIONcompsoc
  \usepackage[nocompress]{cite}
\else
  \usepackage{cite}
\fi

\ifCLASSINFOpdf
\else
\fi

\begin{document}

\title{Adaptive Predictive Power Management for Mobile LTE Devices}

\author{Peter Brand,
  Joachim Falk,
  Jonathan Ah Sue,
  Johannes Brendel,
  Ralph Hasholzner, and\\
  J\"urgen Teich%
\IEEEcompsocitemizethanks{%
\IEEEcompsocthanksitem Peter Brand, Joachim Falk, and J\"urgen Teich are with the Friedrich-Alexander-Universit\"at Erlangen-N\"urnberg\hfill
  $\lbrace$firstname.lastname$\rbrace$@fau.de\protect\\
\IEEEcompsocthanksitem Jonathan Ah Sue, Johannes Brendel, and Ralph Hasholzner are with Intel Deutschland GmbH\hfill
  $\lbrace$firstname.lastname$\rbrace$@intel.com}%
\thanks{}}

\markboth{}%
{Brand \MakeLowercase{\textit{et al.}}: Adaptive Predictive Power Management \\for Mobile LTE Devices}

\IEEEtitleabstractindextext{%
\begin{abstract}
Reducing the energy consumption of mobile phones is a crucial design goal for cellular modem solutions for LTE and 5G standards.
In addition to improving the power efficiency of components through structural and technological advances, optimizing the energy efficiency through improved dynamic power management is an integral part in contemporary hardware design.
Most techniques targeting mobile devices proposed so far, however, are purely reactive in powering down and up system components.
Promising approaches extend this, by predicting and using information from the environment and the communication protocol to take proactive decisions. 
In this paper, we propose and compare two proactive algorithmic approaches for light-weight machine learning to predict the control information needed to allow a mobile device to go to sleep states more often, e.g., in time slots of transmission inactivity in a cell.  
The first approach is based on supervised learning, the second one based on reinforcement learning.
As the implementation of learning techniques also creates energy and resource costs, both approaches are carefully evaluated not only in terms of prediction accuracy, but also overall energy savings.
Using the presented technique, we observe achievable energy savings of up to 17\%.

\end{abstract}

}

\maketitle

\IEEEdisplaynontitleabstractindextext

\IEEEpeerreviewmaketitle

\IEEEraisesectionheading{\section{Introduction}\label{sec:introduction}}
Regarding usability and marketability of battery-powered consumer products such as mobile phones, the minimization of energy consumption in operation is a must.
This is especially important, as battery technology is not advancing fast enough to cope with increased energy consumption of, e.g., displays or application processors.
Depending on various factors, the modem (as the focus of this paper) can be a major contributor to overall energy consumption being responsible for up to 65\,\% of the overall consumed energy~\cite{6951904}.
Apart from structural and technological improvements aiming to lower the power consumption of integrated circuits regardless of their operational mode, \ac{DPM}~\cite{benini2000survey} serves to reduce energy consumption by switching idle components from high power to less power-consuming modes.
This is realized via so-called \emph{policies} that represent schedules of power states for all components of a system.

In \ac{LTE}~\cite{sesia2011lte} and 5G, base stations are responsible for scheduling the traffic in a cell to and from all mobile devices.
While this scheduling is entirely known to the base station, being the communication master, the mobile devices have no knowledge when they will be granted time slots for transmission (via so-called \emph{grants} in the control channel).
In order to guarantee that each such grant is received, each device would need to continuously monitor the control channel.
Due to factors like radio quality or other mobile devices in the cell, there can be a significant amount of time and energy spent by a modem to decode the control channel only to realize that there was no relevant grant transmitted.

Therefore, we propose proactive \ac{DPM}-techniques based on machine learning that predict whether a mobile device will receive a relevant grant at a certain time step.
Contrary to reactive \ac{DPM} techniques that take decisions only after perceiving a grant presence or absence, our presented proactive approach monitors the control channel only when a grant presence is predicted.
Whenever a grant absence is predicted, distinct components of the modem can be put to low power states significantly sooner.

Because of the possibility of a high variability in the scheduling behavior of a base station, we argue that proactive \ac{DPM} needs to be capable of being trained online without any prior experience in a cell. 
In this context, we propose and compare two algorithmic approaches for light-weight machine learning to predict the control information needed to allow a mobile device to save energy, e.g., in time slots of transmission inactivity in a cell.  

The first approach is based on supervised learning, the second one based on reinforcement learning.
As the implementation of such learning techniques also creates energy and resource costs, both approaches are carefully tuned and evaluated not only in terms of prediction accuracy, but also in terms of overall energy savings, based on both simulated data as well as data captured in real LTE cells.
From the viewpoint of a mobile device in an \acs{LTE} environment, a computationally light-weight adaptive predictive \ac{DPM} system that is able to learn online to achieve net energy savings is therefore the preferred solution.

The rest of the paper is structured as follows:
First, we discuss related work in~\cref{sec:relwork}.
Next, in~\cref{lteoverview}, we introduce the basics of the \ac{LTE} communication protocol.
Here, we introduce which control events are relevant for adaptive \ac{DPM}.
Subsequently, in~\cref{sec:arch}, we discuss the architecture of our LTE modem as well as the different power saving policies that can be realized by this architecture.
In~\cref{formal}, we propose two techniques for predictive power management based on machine learning.
Here, a special focus is put on data formats, input and output, and the definition of policies.
Next, in~\cref{pred}, we discuss the applicability and advantages of different machine learning approaches, including their capabilities for online learning and giving a definition of classification errors.
Moreover, we formalize the prediction problem both as a reinforcement learning problem and formulate a supervised learning problem.
Based on this, a quantitative study of computational complexity as well as resource and energy requirements for the implementation of the discussed algorithms is given in~\cref{experiments}.
Experiments are conducted to show that significant net energy savings are achievable in the steady state of scenarios such as video download.
Finally, we conclude with an outlook on future research directions in~\cref{conclusion}.

\section{Related Work}\label{sec:relwork}
A comprehensive overview of \ac{DPM} is provided in \cite{benini2000survey}.
Here, \ac{DPM} techniques are grouped into (i) \emph{adaptive predictive} and (ii) \emph{stochastic} approaches.

Stochastic control problems are characterized by their power state transition times being non-deterministic and more than two power states (i.e., not only on/off) available to components. 
Most related and relevant to our work are \emph{adaptive predictive approaches}:
Predictive power management~\cite{Srivastava:1996:PSS:231913.231922,Chung:1999:DPM:339492.340026, mannor2006machine, Wang:2011:DNP:2024724.2024735} uses machine learning techniques to predict, e.g., the length of idle intervals.
This information can be used to choose an optimal timeout policy or to define suitable power management policies of when to power down/up individual components.
Workload-dependent \acp{DPM} for, e.g, multi-processor systems \cite{Dhiman:2006:DPM:1233501.1233656, Wang:2011:DNP:2024724.2024735}, often act event-triggered.
Here, the DPM system receives a signal indicating that a component is idle and only then has to perform a prediction of an optimal time-out length. 

In contrast to this, in the LTE context and for mobile devices, suitable control signals have to be determined first to trigger dynamic power management actions. Second, in order to save considerable amounts of energy, such control signals themselves must be predicted. Finally, instead of event-triggering, we rather encounter a periodic, time-triggered control problem.

In this paper, we base our modeling of the \ac{LTE} environment and \ac{LTE}-compliant behavior on \cite{sesia2011lte} and \cite{3gppqos}.
For LTE or wireless networks in general, there exist only very few approaches to predictive \ac{DPM} that can be distinguished by the location where predictions are performed: (i) the core network or (ii) the individual mobile device.

In case of the core network, research work has focused on the question of how to save power by optimizing resource block allocation, see e.g. \cite{xu2017deep, bhaumik2012cloudiq, santos2017simple}.
These works aim to reduce power consumption of the average power of all devices registered in a cell by optimizing scheduling decisions from the base station through machine learning.
In contrast to this, our approach is employed on the side of the mobile device in such a network, unaware of the often dynamic scheduling policy of a base station and without any global knowledge on the number of devices in a cell and the number and types of current requests.

Two main challenges have to be coped with in this context: Scarcity of global knowledge available to the predictor in the device as well as scarcity of computational resources available in the modem.
A further difference is the strict requirement of reduced computational complexity for our use-case, as a mobile device is both battery-powered and severely constrained in computational power.

Despite these huge challenges, both approaches might be applied hand-in-hand, as our approach described in the following does not assume any knowledge given on the scheduling techniques used in a base station. 
 
Finally, there also exist a few approaches on adaptive power management on mobile devices that can be distinguished by their adaptiveness to new situations, e.g., number of devices, in a cell and the required additional interaction with the core network.
For example, the work in \cite{Kravets:2000:APM:352157.352160} examines how mobile device receiver behavior can be adapted to increase energy efficiency.
However, the proposed solution requires to extend the protocol by additional signals that indicate superfluous computations in order to notify the system in advance of opportunities to switch components to low-power states.
Although requiring no forecast, this technique would only be applicable in case the full LTE communication protocol would be extended by the proposed signaling mechanism.
The authors of \cite{deng2012traffic} use machine learning to predict when data transmissions are sparse enough to be delayed without the user noticing.
Upon prediction of such an opportunity, the mobile device then signals to the base station to defer the communication, leading to more bursts in traffic, reducing the amount of time data has to be received.
In contrast, our approach proposed in this paper works on top of any MAC layer communication protocol without any change, and requiring no additional control signals.

In \cite{brand2017exploiting}, an approach for grant prediction in LTE is introduced and shown in theory to be beneficial for predictive \ac{DPM}, but no actual prediction technique is proposed and only ideal energy savings are reported.
Additionally to \cite{brand2018reinforcement}, where a solution to the prediction problem is proposed, this paper presents an in-depth investigation of the impact of trace characteristics on theoretically as well as realistically achievable energy savings for a representative application model from \cite{ameigeiras2012analysis}.
Finally, \cite{sue2016binary} presents an approach to the problem of grant prediction based on a supervised learning with impressive false negative prediction rates of only about $2\,\%$.
However, all training there is performed offline on pre-recorded communication traces as presented in \cite{sue2018predictive} and without any online validation or training, thus neglecting the dynamic behavior in a cell caused by changes in the number of active mobile devices and their requests as well as adaptive scheduling behavior of the corresponding base station. 
As a result, no margins of achievable net energy savings for real environments are known.
To the best of our knowledge, this is the first paper to (i) present and compare different approaches for LTE grant prediction for \ac{DPM} in mobile devices that (ii) may be applied in any LTE network without (iii) any prior training, and (iv) analyzes net energy savings in (v) real environments on the basis of sound power and energy models of modem and predictor components.

\section{LTE Protocol}\label{lteoverview}

This section presents an overview of the key control signals of \ac{LTE}. 
With an appropriate level of abstraction, we introduce (i) key layers in LTE communication, (ii) radio transmission and reception concepts, and finally (iii) both uplink and downlink control signals.
Later sections will use these signals to motivate \ac{DPM} strategies and design predictors to realize these strategies.
For a more in-depth and complete \ac{LTE} overview, we refer the reader to \cite{sesia2011lte}.

\subsection{LTE Base Terminology}\label{ltebackground}
\begin{figure}[b]
  \centering
  \includegraphics[width = .47\textwidth]{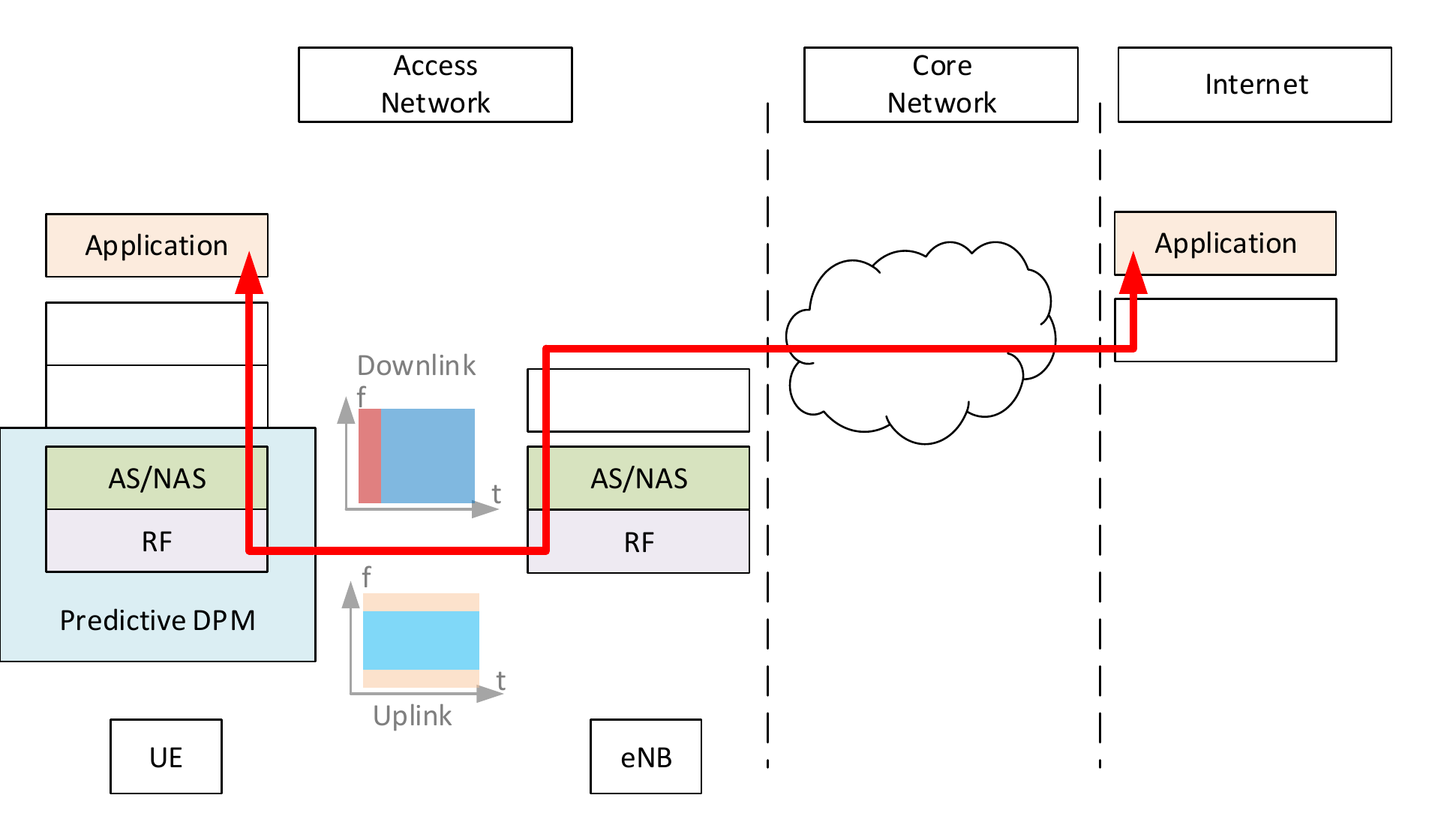}
  \caption{\label{fig::eutran}Overview of an LTE network with the functional division into access network with \acp{UE} and \acp{eNodeB}, here only hinted at core network, and a service provider, i.e., an internet server.
    The red arrow connects two communicating applications and shows how data is sent and received through the protocol stacks.
    For the sake of brevity, we only highlight those protocol layers that are directly responsible for the \ac{RF} traffic between \acp{UE} and \acp{eNodeB}.
    For more details, see~\cite{sesia2011lte}. }
\end{figure}

An \ac{LTE} network consists of two major parts: (i) the \emph{access network} and (ii) the \emph{core network}.
The access network is a cell-structured network of \acp{eNodeB} that communicate with mobile devices (i.e., \acp{UE}) via radio transmission.
As the \ac{UE} modem components -- responsible for radio transmission and reception -- are the focus of this work, our explanation of the \ac{LTE} protocol layers will be focused on the layers that either realize or directly affect the communication between the \acp{UE} and \acp{eNodeB}.
This abstract network view is shown in~\cref{fig::eutran} with the relevant layers -- application layer, \ac{AS}/\ac{NAS} layers, and \ac{RF} layer -- highlighted.
\begin{enumerate}
  \item[(i)] The application layer executes user applications, e.g., video streaming, that connect a \ac{UE} with a service provider, like an internet server, through the core network.
    A functional model of such communication on application level is given in \cref{subsec::funcmodel}.
  \item[(ii)] The \ac{AS}/\ac{NAS} layers generate and process \ac{LTE}-conform control signals as outlined in~\cref{sec::downlink,sec::uplink}.
  \item[(iii)] The \ac{RF} layer realizes the radio signal transmission and reception.
    Analog radio signals are sent between \ac{UE} and \ac{eNodeB} aligned to channels divided in frequency into uplink and downlink regions as detailed in~\cref{sec:comchan}.
\end{enumerate}

\subsection{Communication Channels}\label{sec:comchan}
\begin{figure}[t]
\centering
\includegraphics[width = .45\textwidth]{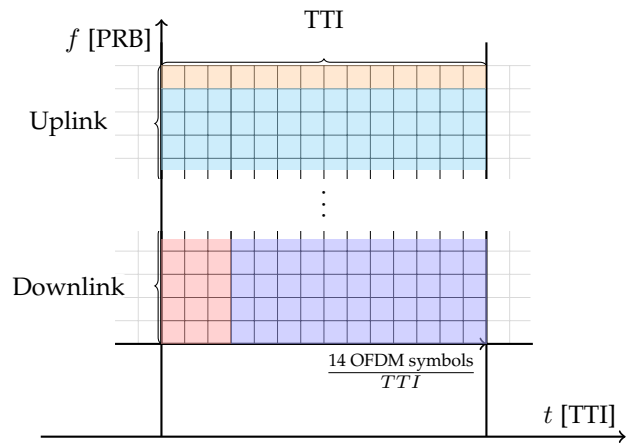}
\caption{\ac{LTE} communication. Depicted is one \acf{TTI}) which has the duration of 1 ms comprising 14 \ac{OFDM} symbols, subdivided in frequency into \acfp{PRB}. }\label{fig::tti}
\end{figure}

\ac{RF} communication in \ac{LTE} between a \ac{UE} and an \ac{eNodeB} is aligned in time to so-called \acp{TTI}, as shown in~\cref{fig::tti}, which are $1$\,ms in length.
Moreover, each \ac{TTI} itself is divided into $14$ \ac{OFDM} symbols of equal length, each of length $\frac{1}{14}$\,ms.
In the frequency domain, all communication is aligned to so-called \acp{PRB} of size $180$ kHz.
Furthermore, downlink (sent from the \ac{eNodeB} to the \ac{UE}) and uplink (sent from the \ac{UE} to the \ac{eNodeB}) are separated in frequency.

To make sure transmissions are free of interference, an \ac{eNodeB} schedules all uplink and downlink data in frequency and time and sends this schedule information to all \acp{UE}.
As the target of the discussed predictive \ac{DPM} (see~\cref{powerman}) are the components of the modem radio frequency reception chain, an in-depth understanding of the downlink (\cref{sec::downlink}) and uplink (\cref{sec::uplink}) region is necessary.

\subsection{Downlink}\label{sec::downlink}

\begin{figure}[t]
\centering
\includegraphics[width = .35\textwidth]{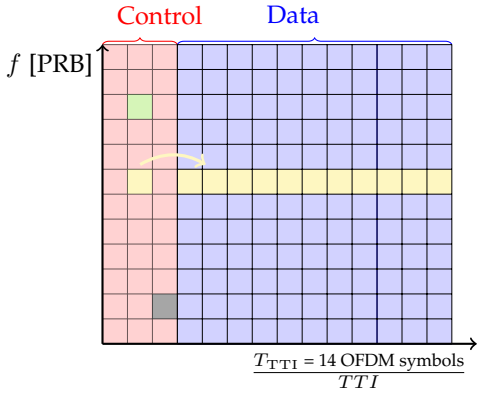}
\caption{Visualization of the downlink region divided into (i) control and (ii) data channel. 
A downlink grant (yellow) always contains information in the control and the data channel.
As can be seen, a downlink grant is always directly followed by a data transmission in the subsequent data channel. Depicted also is an uplink grant (green) and an ACK/NACK signal (black) with information only contained in the control channel. }\label{fig::dlTti}
\end{figure}

The downlink region of a \ac{TTI} is divided (in time) into the \emph{control} channel (first 3 \ac{OFDM}-Symbols) and the \emph{data} channel (remaining 11 \ac{OFDM}-Symbols), see~\cref{fig::dlTti}.
As is implied by the names, the data channel is reserved for actual user data, while the control channel holds all relevant protocol information.

\cref{powerman} will show how a prediction of the information contained here can be leveraged to obtain energy savings compared to a state-of-the-art \ac{DPM}.

\subsubsection{Grant Signaling}
Grant signaling is used to notify the \acp{UE} of scheduling opportunities with (i) \acp{DLG} and (ii) \acp{ULG}.
These grants are used for the purpose of notifying a \ac{UE} of specific \acp{PRB} at specific times where (i) downlink transmissions are to be received or where (ii) uplink transmissions are allowed.
Each type of grant information is signaled in the control channel of a \ac{TTI} and addressed to a specific \ac{UE}.
Moreover, downlink grants always point to \acp{PRB} in the data channel of the very same \ac{TTI}, whereas uplink grants indicate \acp{PRB} to be used by the \ac{UE} exactly $4$ \acp{TTI} in the future.
Finally, to serve as input for our work on grant prediction in~\cref{pred}, downlink and uplink grants can be formally described by tuples:
\begin{equation}\label{eq::dlg}
DLG = (ndi, tbs, mcs) \in \mathbb{B} \times \mathbb{N}^+\times \mathbb{N}^{[0,31]} 
\end{equation}
\begin{equation}\label{eq::ulg}
ULG = (ndi, tbs, mcs) \in \mathbb{B} \times \mathbb{N}^+ \times \mathbb{N}^{[0,31]}
\end{equation}
$ndi$ is a Boolean value that describes whether a new packet ($ndi \equiv \true$) should be sent, or a previous packet should be retransmitted ($ndi \equiv \false$).
$tbs$ is a positive natural number that specifies the exact number of \acp{PRB} associated with the grant within the data channel. 
$mcs$ is also a  natural number between $0$ and $31$ that specifies the \ac{MCS} for both uplink transmission and downlink reception and describes how the RF payload data is encoded.

\subsection{Uplink}\label{sec::uplink}
Not crucial for prediction, but nevertheless important for our evaluation of overall energy consumption in \cref{experiments}, are the uplink RF \ac{LTE} mechanisms.
In this work, we consider (i) \acp{BSR}, (ii) uplink data transmissions, and (iii) ACK/NACK signaling.
\begin{itemize}
  \item[(i)] \acp{BSR} are sent by the \ac{UE} to notify the \ac{eNodeB} of data that is ready to be sent.
    Hence, asking for future scheduling and, thus, an uplink grant.
  \item[(ii)] Uplink data transmission is implicitly scheduled exactly 4 \acp{TTI} after the reception of an uplink grant.
  \item[(iii)] ACK/NACK signaling in the uplink is realized to notify the \ac{eNodeB} of successful (ACK) or unsuccessful (NACK) data reception from the \ac{UE}, and is specified to happen exactly 4 TTI after downlink transmissions (i.e., a received DLG).
\end{itemize}
Since the \ac{eNodeB} knows when in time and where in frequency to expect data transmissions and ACK/NACK feedback, it will react to unexpected behavior (no, or erroneous data transmissions, missing ACK/NACK feedback).
This is done by rescheduling and resending an uplink grant (missing transmission) or resending the downlink data (missing ACK/NACK), but both kinds of grants with the field $ndi = \false$.

\begin{figure}[t]
\centering
\includegraphics[width = .43\textwidth]{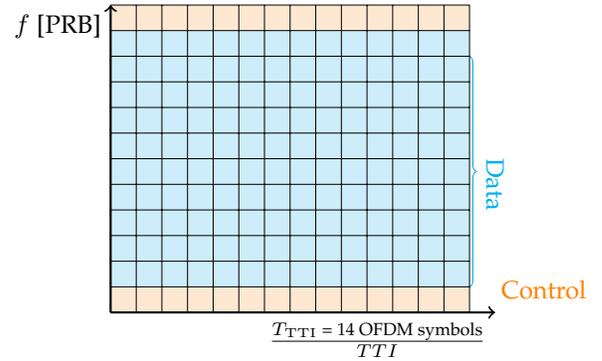}
\caption{Visualization of the uplink region. The upper- and lowermost frequency bands (the uplink control channel) are exclusively used for protocol signaling. The uplink data channel is reserved for sending payload. If a \ac{UE} transmits both control signals and data, both is bundled together and sent within the data channel.}\label{fig::ulTti}
\end{figure}

Similar to the downlink case, the uplink frequency spectrum is subdivided into channels for uplink control and for uplink data, as shown in~\cref{fig::ulTti}.
Here, the upper- and lowermost \acp{PRB} belong to the control channel and the frequencies in between carry the data channel.

Transmissions in a \ac{TTI} are exclusive to one channel, which means that a \ac{UE} can either send in the control or data channel, with the data channel taking precedence.
Thus, if a \ac{UE} transmits payload data, everything will be transmitted in bulk in the data channel.
Otherwise, it will be sent in the control channel.

\section{Architecture and Power Modeling}\label{sec:arch}
\begin{figure}[htb]
  \centering
  \includegraphics[width = .45\textwidth]{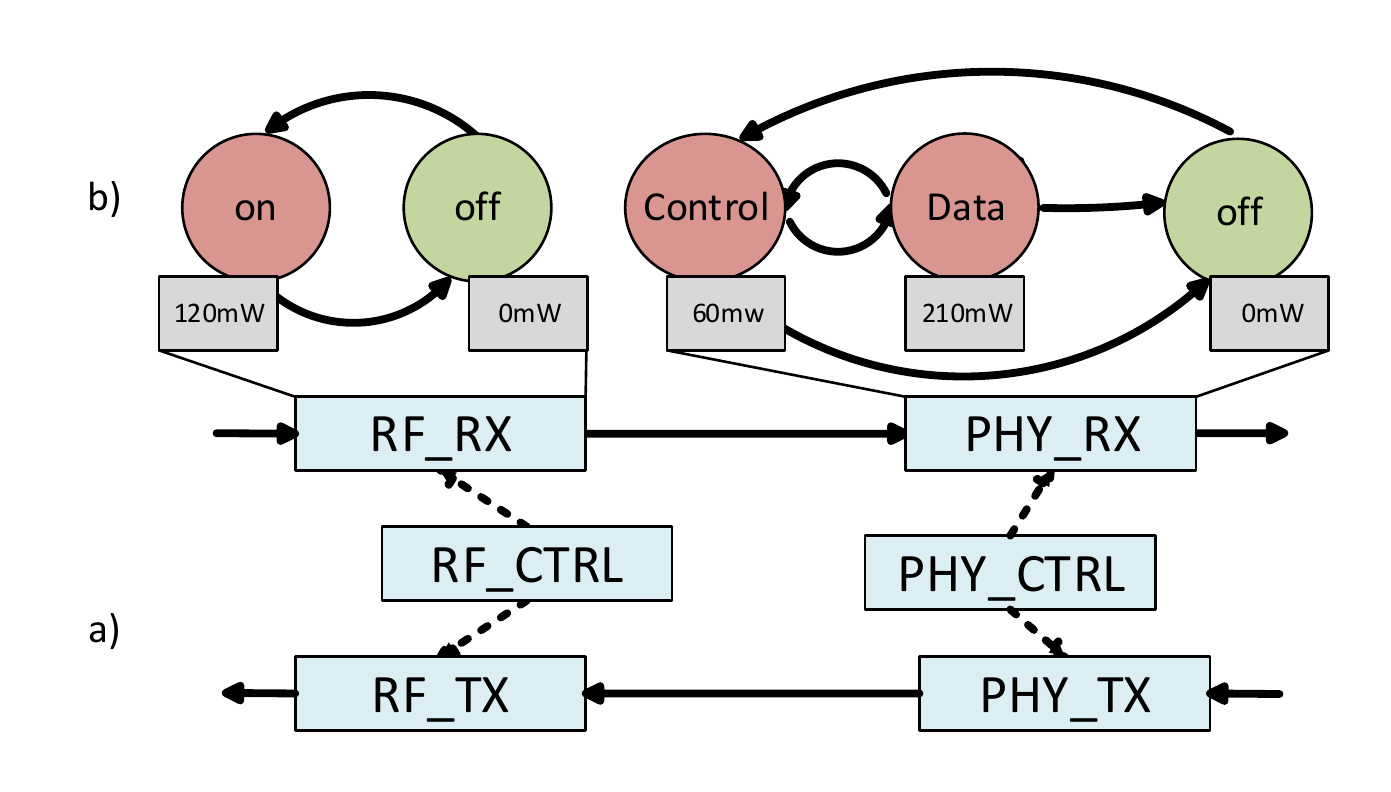}
  \caption{\label{hwandpsm}a) Abstract hardware model of an LTE modem.
    b) Power State Machines (PSMs) \cite{rgtwxh_2014-maestro},\cite{pomacs} of the RF\_RX and PHY\_RX components.
    The power states "on", Control", and "Data" are high, while "off" are low-power states. 
    We assume the transitions to be instantaneous and requiring no additional power.}
\end{figure}

To model the power consumption of a modem and to formalize policies for DPM, we take a \emph{\ac{PSM}-based} approach according to~\cite{6951904}, see~\cref{hwandpsm} for convenience.
Here, each essential power-manageable hardware component is modeled by a PSM with its respective power states and possible state transitions corresponding to power management decisions. Distinguished in~\cref{hwandpsm} are a \emph{Radio Frequency (RF)} part and a \emph{Physical Layer (PHY)} part.
In each power state, a fixed nominal power consumption value is assumed.
Additionally, each part is divided into a reception (RX), transmission (TX), and control (CTRL) components.
Thus, resulting in the six different components shown in~\cref{hwandpsm}a).
The RF components receive information from, respectively, send information to the PHY components that decode, respectively, encode the information.

In the following section, we motivate the potential of power and energy savings in an LTE modem through proactive power management.
We propose DPM policies based on the prediction of grant signals in the reception path, thus affecting the RF\_RX and PHY\_RX components.
Their PSM models are shown in~\cref{hwandpsm}b).
As can be seen, both components can be in a low (off) or high power state (on, Control/Data).
Whereas during the decoding of the control region (first three OFDM symbols of a TTI), the PHY\_RX component is in the \emph{Control} state, it transits to the \emph{Data} state when decoding the data channel.

\subsection{Power Management Policies}\label{powerman}
In the context of an LTE modem, a power management policy denotes a schedule of the power states of all its DPM-controllable components over the duration of a TTI.

\begin{figure}[htb]
  \centering
  \includegraphics[width = .48\textwidth]{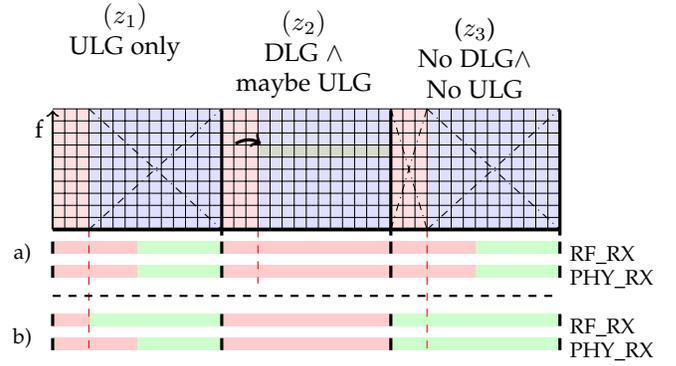}
  \caption{\label{scenarios}Three scenarios and corresponding policies of a) reactive DPM and b) proposed predictive DPM are shown (red/green: component turned on/off).
    In the left scenario $z_1$, an uplink grant (ULG) is received.
    Scenario $z_2$ is characterized by a downlink grant (DLG) and potentially additionally an ULG being received.
    Finally, scenario $z_3$ represents the case of a TTI in which the UE does neither receive a ULG nor DLG.
    As can be seen, the predictive DPM may power the two components RF\_RX and PHY\_RX down earlier and, thus, for a longer duration of time.
    In the right scenario, both components can be turned off even from the very beginning of the TTI.
  }
\end{figure}

On the receiver (RX) path, an \ac{LTE}-compliant reactive DPM could distinguish a very simple policy as shown in~\cref{scenarios}a):
If the data channel contains no data for the UE, which is the case if no DL grant (DLG) is present in the control channel, then the RF\_RX and PHY\_RX components can be simply turned off for the remainder of the TTI, see the two scenarios $z_1$ and $z_3$, respectively, on the left (only an ULG) and on the right (neither ULG nor DLG present) in~\cref{scenarios}.
Otherwise, the components must remain in the high power state and no power savings are possible in this case (scenario $z_2$ shown in the middle).

However, the potential to achieve considerable power and energy savings through reactive DPM in scenarios $z_1$ and $z_3$ is not very high, as the information to power down RX components becomes available to the DPM only after the complete decoding of the data channel  as shown by the length of the red time intervals of the components RF\_RX and PHY\_RX in~\cref{scenarios}a).

In contrast, imagine the DPM could correctly predict the information that neither a DLG nor a ULG grant will be received in the next TTI.
Obviously, both components could then stay in the off mode for the complete duration of a TTI (see scenario $z_3$ in~\cref{scenarios}b).
Moreover, in scenario $z_1$, this would allow at least the RF\_RX component to stay off for a considerably longer time than in case of a reactive DPM.

Our predictive \ac{DPM} approach proposed in this paper aims to exploit the full potential of energy savings of a UE by trying to maximize the intervals (green in~\cref{scenarios}b) in which components are operated in off mode.
This is achieved by aggressively turning the modem components off \textit{before the decoding of the control channel}.
To achieve this, \cref{pred} will introduce and evaluate two machine learning algorithms for \textit{time-series prediction} of whether a TTI will contain a DLG or ULG grant for a UE in the next TTI.
In contrast to a reactive approach, it will be shown that this has the potential of a significant amount of net energy savings (see~\cref{experiments}).
However, prediction bears also the danger of potentially loosing transmission capacity in uplink or downlink in case of misprediction of grant signals.
Such mispredictions may indeed affect performance but notably also future traffic of the whole cell due to retransmissions.
Because the \ac{eNodeB} will generally retransmit data if a UE does not react as expected, e.g., by missing grant signaling due to erroneously turning the modem off, a prolonged data transfer for the affected data will result.
In consequence, no energy savings, but rather increased energy might be observed due to the required retransmissions.
This and the overhead of energy consumed by an implementation of the prediction technique itself must therefore be carefully evaluated.

\section{Predictive DPM}\label{formal}

This section formalizes the problem of predictive DPM, introduces the notion of DPM policies based on the three scenarios distinguished previously, and provides formulas for energy estimation based on the notion of analyzed traces.

\begin{figure}[t]
  \centering
  \includegraphics[width = .48\textwidth]{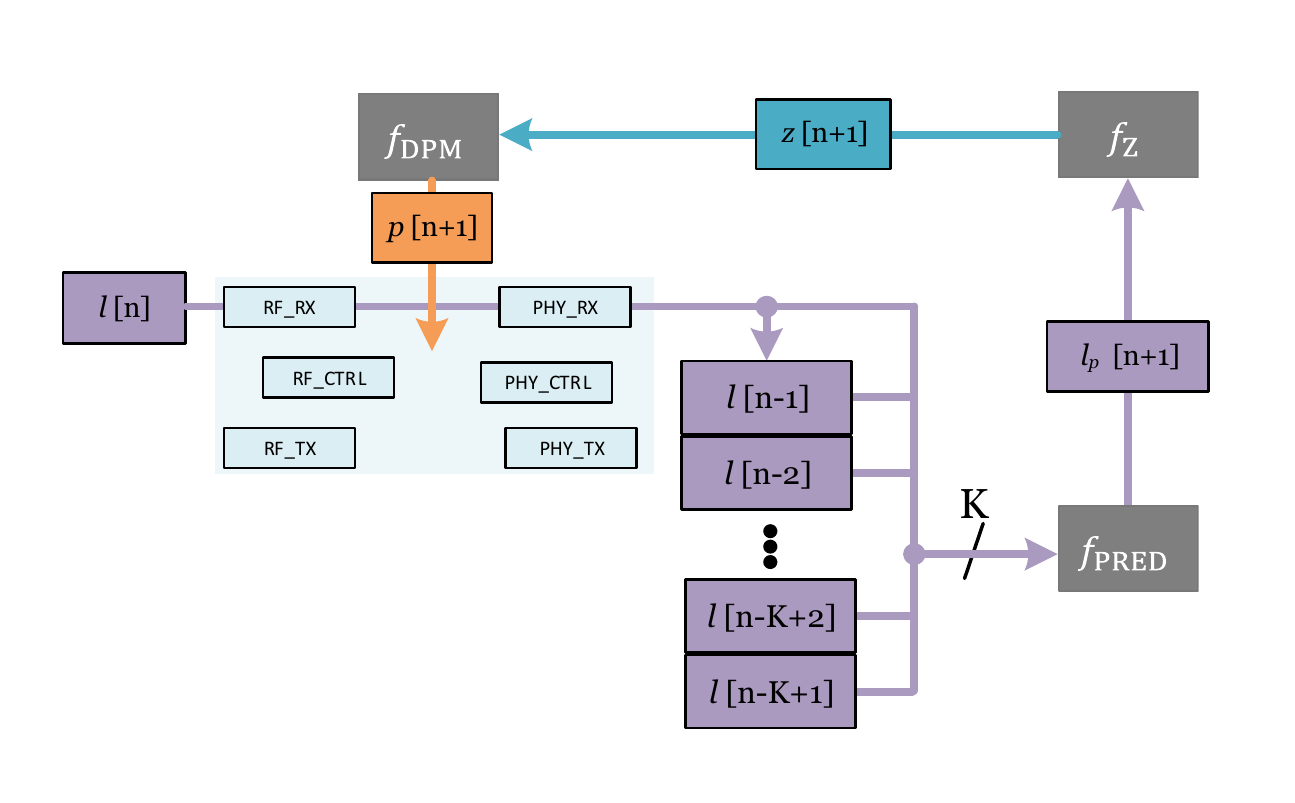}
  \caption{\label{overview}Predictive LTE-modem DPM system.
    It consists of a predictor that computes a prediction $\vec{l}_p[n+1]$ of the next TTI's control information (based on a sliding windows of the last K information) and a scenario recognizer that maps this prediction to a scenario $\vec{z}[n+1]$ as identified in~\cref{scenarios}.
    From this, an optimized policy $\vec{p}[n+1]$ is derived.
    This policy is immediately applied to all components of the modem at the beginning of the next TTI.}
\end{figure}

\Cref{overview} gives an overview of the complete predictive DPM system.
Based on received LTE traces with $\vec{l}[n]$ denoting the control information received by the modem of a UE in the $n$th TTI of a trace (see~\cref{traffic}), a predictor as explained in~\cref{preddpm} performs the prediction $\vec{l}_p[n+1]$ of the next TTI's control information upon which the next scenario $\vec{z}[n+1]$ is determined.
\Cref{policy} thereby illustrates the notion of policies as schedules of power states of each power-controllable component and how they are determined for each of the characterized scenarios to save the highest amount of energy.

\subsection{LTE Traffic Modeling}\label{traffic}

Let the \emph{TTI information} $\vec{l}[n]$ represent all information that is observable by a single UE during the $n$th TTI.
This information is used for grant prediction.
It is defined as a tuple containing the ULG and DLG information, as introduced in~\cref{eq::dlg,eq::ulg}:
\begin{equation}
  \vec{l}[n] = (\mathrm{ULG}, \mathrm{DLG}) \in L
\end{equation}
Based on this, a  so-called \emph{trace} of length $N$ can be modeled by a sequence
\begin{equation}
  \vec{l} = <\vec{l}[1], \vec{l}[2],...,\vec{l}[N]> \in L^N
\end{equation}

\subsection{Policies and Mapping of Scenarios to Policies}\label{policy}

A \emph{policy} $p$ is defined as a schedule of the power states of each power-controllable component 
$r\in R = \{r_\mathrm{RF\_RX},\allowbreak{} r_\mathrm{PHY\_RX},\allowbreak{} r_\mathrm{RF\_TX},\allowbreak{} r_\mathrm{PHY\_TX},\allowbreak{} r_\mathrm{RF\_CTRL},\allowbreak{} r_\mathrm{PHY\_CTRL}\}$ according to~\cref{hwandpsm} within a \ac{TTI}.

Assuming that the initial state of each component is a high power state and that each component can maximally be switched off once during a TTI, the schedule of each component $r$ may be described by a time index 
\begin{equation}
t_r \in \{ \frac{0}{W},\frac{1}{W},\frac{2}{W}\cdots,\frac{W}{W} \} \cup \{ - \}
\end{equation}
that indicates the fraction of a TTI (with $W=14$) at which the component is powered down to its off-state.
A policy $p$ is then simply a tuple $p = (t_1, \cdots, t_{|R|})$ of such time indices.

Hereby, we assume that an \ac{LTE}-conform DPM system will never switch a component to on, apart from the very beginning of a TTI.
Following this, the time index of $0$ means that the respective component will not be switched on at all, while $1$ means it is kept on during the entire TTI.

As motivated in the previous section (see~\cref{scenarios}), we distinguish three relevant scenarios $Z = \{z_1, z_2, z_3\}$ for our LTE-modem DPM.
In detail, scenario $z_1$ corresponds to information \textit{only} in the control channel, scenario $z_2$ has information \textit{both} in the control and data channel, and scenario $z_3$ indicates no information within the TTI.

The mapping of this input to power states and transitions is intuitive:
For $z_1$, only the control channel should be received, i.e., RX components can be turned off after the control signals have been processed.
For $z_2$, obviously all content of the TTI has to be received, and no RX component should be turned off.
Last, for $z_3$, the RX components can be turned off during the whole TTI without any loss of information.
The time that the RX components need to process the data contained in the control channel and can forward the information to the DPM system is dependent on the modem itself.
In the following energy analysis, we assume the RX components are operated at a 50\,\% duty cycle and turned off in the second half ($0.5\cdot t_\mathrm{TTI}$) of the TTI if no DL grant is present in the control channel.

Obviously, three policies, one characterizing each of the above scenarios are necessary but also sufficient to describe the DPM schedules for each modem component.
\Cref{traces} shows the relation between TTI tuple information $l$, the corresponding scenario $z$, and the resulting policy $p$ to be chosen for the purpose of energy reduction:

\begin{table}[htb]
  \centering
  \begin{tabular}{ccl} 	  
    \toprule
    TTI                             & Scenario & Policy                         \\
    $l$                             &  $z$     & $p$                            \\
    \midrule
    $(\true,\false)$                & $z_1$    & $p_1 = (0.5 , 0.5 , -, -,1,1)$ \\ 
    $(\true,\true), (\false,\true)$ & $z_2$    & $p_2 = (1,1,-,-,1,1)$          \\ 
    $(\false,\false)$               & $z_3$    & $p_3 = (0,0,-,-,1,1)$          \\ 
    \bottomrule
  \end{tabular}
  \vspace{1mm}
  \caption{\label{traces}Mapping of a TTI information tuple $l$ to a scenario $z$ and corresponding DPM policy $p$.
    Each element in the policy vector corresponds to one physical modem component, the value indicating at which portion of TTI length the component is switched to the off state.
    The don't care values $-$ for the transmitter components $r_\mathrm{RF\_TX}$ and $r_\mathrm{PHY\_TX}$ stem from the fact that our proposed policies only target the RX chain of the modem.}
\end{table}

Note that the mapping of TTI information $l$ to a scenario $z$, respectively policy $p$, can also be used in the following predictive approach, thus, deriving $\vec{z}[n+1]$ and policy $\vec{p}[n+1]$ directly from a predicted tuple $\vec{l}_p[n+1]$ according to~\cref{traces}.
Next, we will explain the prediction mechanism itself, which is based on an observed time series of TTI control information.

\subsection{The Predictive DPM Cycle}\label{preddpm}

In contrast to a reactive DPM system that may take as input only TTI control information up to, respectively of the current TTI $n$, a time-series based predictor takes a sliding window of length $K$ to predict the TTI control information $\vec{l}[n+1]$ of the next TTI, see~\cref{overview}.
Such a predictor may be represented by a function $f_\mathrm{PRED}: L^\mathrm{K} \rightarrow L$ that uses the information of $\vec{l}[n-K+1] \ldots \vec{l}[n]$ to  compute a prediction $\vec{l}_p[n+1]$ of the next TTI control information tuple.
As the predictor has to predict whether a grant will be present or not in the next TTI, only the four discrete tuples shown in \cref{traces} will result from prediction, as the predictor will never predict the TTI information value $\unknown$.

Based on the prediction $\vec{l}_p[n+1]$, the scenario $\vec{z}[n+1]$ is determined by the mapping function $f_\mathrm{Z}: L \rightarrow Z$ according to~\cref{traces}.
The table also describes the final step of mapping the scenario $\vec{z}[n+1]$  to a desired policy.
This mapping can therefore also be described by a function $f_\mathrm{DPM}: Z \rightarrow P$ according to~\cref{overview}.

\subsection{Energy Estimation}\label{energyestimation}

Since the motivation for applying learning techniques is to reduce the modem power, respectively energy consumption, it is important to consider the inherent energy consumption of the prediction overhead per TTI as well.
In the following, this overhead is denoted by $E_\mathrm Q$.
This overhead will be carefully examined in our experiments in terms of clock cycles and energy consumption when implementing the predictor in software on a \ac{DSP} that is representative for usage in LTE modems (see \cref{experiments}).
As our goal is to evaluate net energy savings, we take the energy consumption $E^\mathrm{DPM}$ of the reactive approach as a base line for the comparison with the energy consumption $E^\mathrm{COG}$ -- including prediction overhead $E_\mathrm Q$ -- obtained by the predictive DPM approach when analyzing a given trace $\vec{l}$.

The energy consumption $E^\mathrm{COG}$ for a given trace is dependent on the cognitive policy $\vec{p}^\mathrm{COG}[n]$ (with $\vec{p}^\mathrm{COG}[n].t_r$ carrying the respective time portion of TTI $n$ the resource $r$ will stay in its "on" power state) and the power consumption $P^\mathrm{on}_r$ ($P^\mathrm{off}_r$) of each resource $r$ in the high (low) power state.
Thus, the energy consumption of a complete trace using predictive DPM can be calculated as follows:
\begin{align}\label{ecogformula}
   \scalebox{.75}{$E^\mathrm{COG} = \sum\limits_{n = 1}^{N}(E_\mathrm Q$} &\scalebox{.75}{\;+ $\sum\limits_{\substack{r\in R \setminus \\ \{r_\mathrm{RF\_TX},\\r_\mathrm{PHY\_TX}\}}} P^\mathrm{on}_r \cdot \vec{p}^\mathrm{COG}[n].t_r + P^\mathrm{off}_r \cdot (1 - \vec{p}^\mathrm{COG}[n].t_r)$} \nonumber \\
                                                                          &\scalebox{.75}{\;+ $\sum\limits_{\substack{r\in \\ \{r_\mathrm{RF\_TX},\\r_\mathrm{PHY\_TX}\}}} P^\mathrm{on}_r \cdot \vec{p}^\mathrm{DPM}[n].t_r + P^\mathrm{off}_r \cdot (1 - \vec{p}^\mathrm{DPM}[n].t_r))$}
\end{align}

Since the proposed DPM considers only the power state transitions of the RX components, we assume the TX components ($r_2, r_3$) to be steered by a state-of-the-art reactive LTE DPM policy $p^\mathrm{DPM}$.

Likewise, the energy consumption for the reactive approach, dependent on the reactive LTE DPM policy $p^\mathrm{DPM}$, may be computed as:
\begin{align}\label{edpmformula}
  \scalebox{.75}{$E^\mathrm{DPM} = \sum\limits_{n = 1}^{N}\sum\limits_{r\in R} P^\mathrm{on}_r \cdot \vec{p}^\mathrm{DPM}[n].t_r + P^\mathrm{off}_r \cdot (1 - \vec{p}^\mathrm{DPM}[n].t_r)$}
\end{align}

A net energy gain can be achieved if $E^\mathrm{COG} < E^\mathrm{DPM}$ holds for a trace.

\section{Machine Learning Techniques for LTE Grant Prediction}\label{pred}

There exists a multitude of machine learning approaches.
Which of the following approaches is suitable or not depends mainly on which information is available during learning.
If during each learning step, there is direct feedback on the quality of a classification or prediction, \emph{supervised learning} is appropriate.
If this information is not readily available at every training step, but rather an approximation of quality must be deduced indirectly from certain events, \emph{reinforcement learning} may be the appropriate choice.
Finally, \emph{unsupervised learning} may be applied to cases where neither of the above information is available.
In our case of LTE grant prediction, a suitable machine learning approach shall correctly predict the next TTI control information based on a sequence of $K$ previous control tuples.
Thus, since the correctness of each prediction may be asserted either directly or indirectly (listening and grant being sent or not listening when no grant is being sent), we do not consider unsupervised learning as a preferable technique for our problem.

Before delving into details, \cref{class} discusses error classes and how mispredictions may affect a UE and even the whole cell.
Subsequently, in~\cref{online}, both a supervised and a reinforcement learning approach are presented and benefits and shortcomings of each outlined.
Next, we present solutions for LTE grant prediction based on supervised learning (\cref{superv}) and reinforcement learning (\cref{reinfo}).

\subsection{Error Classification}\label{class}
In the context of our stated LTE grant prediction formalization, the scenario identification is dependent on the predicted presence or absence of a DLG and an ULG.
Hence, it can be defined as a binary classification problem for which two types of prediction errors exist: \textit{False Positive (FP)} and \textit{False Negative (FN)} errors~\cite{sue2016binary}.

A FP error means a grant was erroneously predicted to appear.
This error is neutral to performance (e.g., data rate), as no information is lost.
However, as a false positive error means that components are left in a high power state for longer than needed, energy savings are missed, thus, negatively affecting the non-functional property energy consumption.
To measure the proportion of false positive errors, we define the \emph{\acp{FPR}} as:
\begin{equation}\label{fpr}
\text{FPR} = \frac{\#\,\text{grant presences erroneously predicted}}{\#\,\text{grant absences}}
\end{equation}

The second kind of prediction error is a FN error, which means a grant \emph{absence} was erroneously predicted.
As explained in~\cref{ltebackground}, this leads to scheduled \acp{PRB} to be effectively unused because either the downlink data is not received by the \ac{UE} or the \ac{UE} will not transmit any uplink data.
False negative errors are diminishing the performance, i.e., leading to a deterioration of data rate and effective bandwidth of the whole \ac{LTE} cell.
Analogous to the \ac{FPR}, we define the \emph{\ac{FNR}} as:
\begin{equation}\label{fnr}
\text{FNR} = \frac{\#\,\text{grant absences erroneously predicted}}{\#\,\text{grant presences}}
\end{equation}

In summary, minimization of the FPR corresponds to a minimization of energy consumption as defined by~\cref{ecogformula}, while minimization of the FNR is necessary in order not to affect the quality of service.
Hence, the two approaches presented in the following both aim at saving energy while minimizing the FNR.

\subsection{Design Considerations}\label{online}
Depending on \ac{UE} application, UE movement, and other \acp{UE} behavior in the cell, traffic patterns are diverse and may undergo continuous changes.

Thus, when designing a predictor, there are several conflicting design objectives.
The first pair is complexity and prediction accuracy.
As explained in~\cref{energyestimation}, performing a prediction itself as well as the training of the predictor requires additional computations.
Assuming a nominal prediction accuracy, a less complex prediction algorithm will yield higher energy savings compared to a more complex algorithm.
Of course, the prediction accuracy of a more complex algorithm may achieve a better prediction accuracy if the trace characteristics are hard to predict.
For traces with especially simple characteristics, complexity may be wasted.

The second important decision is assumption of stationarity.
Assume a trace that is stationary for a long time.
In this case, training a predictor with a quick convergence to a sufficient solution on a short initial part of interval may be the preferred choice.
However, if the observable traces are subject to continuous changes, a solution that is capable of on-line training to improve its prediction may be superior.

Therefore, we propose and compare two different approaches that carefully balance these considerations.
The first presented approach (\cref{superv}) is based on supervised learning.
Here, we propose a predictor based on a neural network, aimed to be trained quickly until it is turned into exploitation mode.
The second approach (\cref{reinfo}) is based on reinforcement learning.
There, we introduce a light-weight tabular prediction algorithm, which is trained continuously.
Finally, in~\cref{experiments}, both approaches are compared for a number of different traces in terms of prediction accuracy and the potential for energy savings.

\subsection{Supervised Learning}\label{superv}

A first benchmark of several supervised ML algorithms for LTE grant prediction is described in \cite{sue2016binary}.
In essence, 3 different algorithms have been trained to perform grant prediction.
Their output, a value between 0 and 1, is used as cost-sensitive classification input, to decide whether a grant, modeled as 1, or no grant, modeled as 0, should be predicted.
This gives the ability to tune the false negative rate (FNR) separately from the accuracy result.
Using this two-stage classification approach has 2 main advantages:
\begin{itemize}
\item \textbf{Intrinsic algorithm accuracy comparison:} Applying the same cost-sensitive classification technique with different first-stage ML algorithm makes it possible to compare the intrinsic predictive ability of these algorithms.
  In other words, it becomes possible to answer the following question: which algorithm can inherently model the grant traffic better?
\item \textbf{Tunable operating point:} For a given prediction, it might be requested to reach different FNR performances.
  For instance, if too many grants have been missed (high FNR) at the beginning of an LTE scenario, it is desirable to adjust only the cost-sensitive parameters in order to have a predictor which is less prone to false negatives than false positives.
  Using such dynamic and straight-forward adjustments of the predictor output can be used to tune the FNR depending on the user needs.
  Therefore, a safe cost-sensitive setup, i.e., low FNR but higher FPR, would be used for scenarios with time constraints where packet losses could be very damageable.
  For non-critical scenarios, a more aggressive cost-sensitive setup, i.e., low FPR but higher FNR, would be applicable.
\end{itemize} 

\subsubsection{Feed-Forward Neural Network}

From 3 popular supervised learning approaches, i.e., feed-forward neural networks (FFNN), support vector regression (SVR) and recurrent neural networks (RNN), we chose the less computationally expensive one, the FFNN approach.
Indeed, SVR requires to solve a high-dimensional optimization problem, often in the dual space \cite{boser1992training}.
RNNs, when unfolded through time, are also computationally very expensive \cite{pascanu2013difficulty}.

\begin{figure}[tb]
  \centering
  \vspace{-3mm}
  \resizebox{0.45\textwidth}{!}{\includegraphics{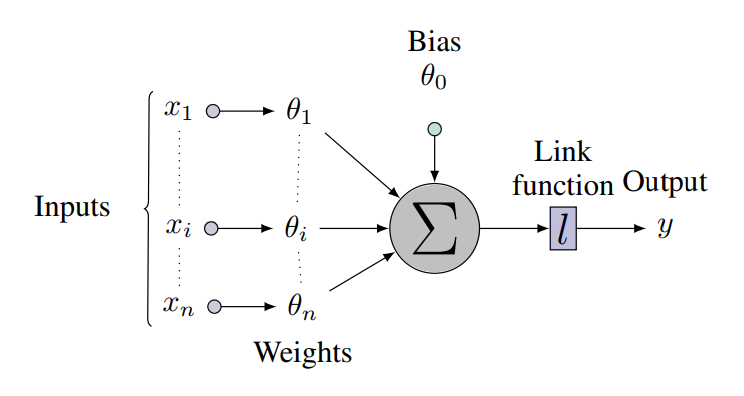}}
  \caption{\label{neuron}Process in one neuron as depicted in~\cref{neu}.
    $\mathbf{\theta}$ is the weight vector subject to backpropagation.
    The link function, $l$, is often sigmoid, linear or tangent hyperbolic.}
\end{figure}

\begin{figure}[tb]
  \centering
  \resizebox{0.45\textwidth}{!}{\includegraphics{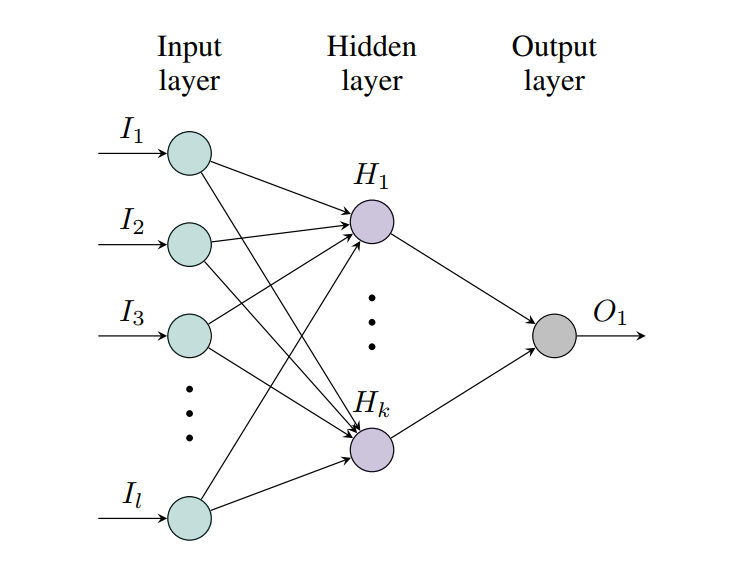}}
  \caption{\label{net}Example of a FFNN with one hidden layer.
    The neurons of the input layer do not act as other neurons as depicted in~\cref{neuron}.
    They store the values presented as input.}
\end{figure}

In particular, we use a specific type of FFNN where all the neurons of one layer are connected to all neurons of the next layer, i.e., fully connected FFNN.
In~\cref{neuron}, we describe the model of one neuron which is the core building block of the entire neural network depicted in~\cref{net}, which uses the hyperparameters given in~\cref{tabParam}.
\begin{table}[bt]
  \caption{\label{tabParam}Neural Network Hyperparameters}
  \centering
  {\setlength{\tabcolsep}{1.5em}
    \begin{tabular}{|l|}
    \hline
    \textbf{Name [properties]}\\
    \hline
    Layer 1 [60 neurons, tangent hyperbolic]\\
    \hline
    Layer 2 [10 neurons, tangent hyperbolic]\\
    \hline
    Output Layer [2 neurons, linear]\\
    \hline
    Performance Function [mean square error]\\
    \hline
    \end{tabular}
  }
\end{table}
Formally, the output of one neuron can be expressed as
\begin{equation}
l(\Sigma)=l(\theta_{0}+\theta_{1}x_{1}+...+\theta_{n}x_{n})\label{neu}
\end{equation}
with $(x_{1},...,x_{n})$ being the output of the $n$ previous neurons, $l$ the link function and $\boldsymbol{\theta} \in \mathbb{R}^{n+1}$ the weight vector modified by backpropagation in order to minimize the error between outputs and targets during the training phase.

This type of ML algorithms is known to be universal function approximators and can therefore achieve any kind of nonlinear mapping between inputs and outputs \cite{cybenko1989approximation}.
Although training neural networks with backpropagation can be computationally expensive, several variants exist~\cite{lecun2012efficient}, allowing to choose the optimal accuracy-complexity trade-off depending on the task.
Generally, no conclusions can be drawn on the influence of the number of hidden layers and neurons.
Therefore, we chose the hyperparameters given in~\cref{tabParam}.

The prediction window input length is chosen equal to $K = 10$, which is equivalent to taking the last 10 values of each LTE metric as neural network input.

In total, the input vector is thus a 60-dimensional vector containing the normalized LTE metric values, explicitly considering all information contained in a grant ($\mathrm{ndi}$, $\mathrm{tbs}$, and $\mathrm{mcs}$).
These values might give some indications on the past grant history but also bandwidth occupancy, channel conditions, and past corrupted data, which are informations used by the eNodeB to decide on the future allocations. 

The output is a real-valued estimation of the likelihood of a ULG and/or a DLG presence in the next TTI.
This real-valued estimation is then transformed by the cost-sensitive classification decision stage to discern the actual predicted TTI (present or absent).

\subsubsection{Cost-Sensitive Classification}

\begin{figure}
  \centering
  \includegraphics[width=250px]{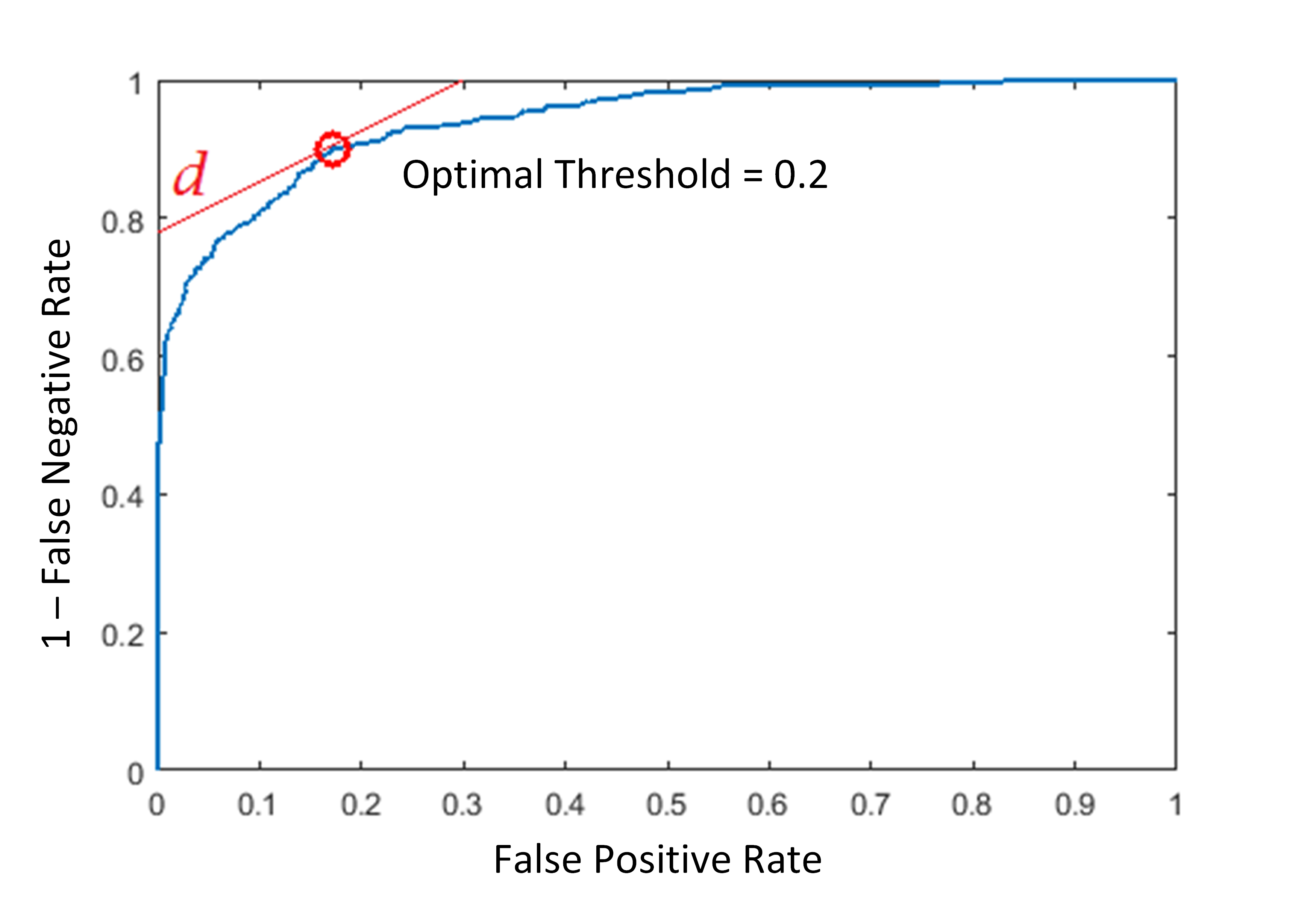}
  \caption{\label{rocfig}A receiver operating characteristic (ROC) representation of the cost-sensitive classification CSC.
    The optimal threshold is obtained by setting $m$, the slope of $(d)$.
    Therefore, the ratio of the \ac{FNR} and \ac{FPR} can be tuned as depicted in~\cref{ratio}.}
\end{figure}

A receiver operating characteristic (ROC) curve is often used to assess the classifiers' performance.
It depicts the trade-off between the \ac{FNR} and the \ac{FPR}.
The performance of a classifier is assessed by computing the area under the ROC curve (AUC).
The higher is the AUC, the more efficient is the classifier.
As depicted in~\cref{rocfig}, the FNR and FPR can be tuned to achieve the any desirable trade-off on the training data.

In the following, let $P(y/t)$ denote the conditional probability that $y$ is predicted given $t$ as target, $P(y,t)$ denotes the joint probability and $C(y,t)$ the cost of predicting $y$ with target $t$.
Therefore, a general cost function $R$ can be defined for the classifier,

\begin{equation}
R = \sum_{t,y = 0,1} P(y,t)\ C(y,t)
\label{costroc}
\end{equation}

In this work, since correct classifications are not penalized, only cases with $C(1,1)=C(0,0)=0$ are considered and from Bayes' rule, joint probabilities can be expressed with conditional probabilities.
Under the naive Bayes assumption, $\textit{FPR}=P(1/0)$ and $\textit{FNR}=P(0/1)$ and therefore the formulation for~\cref{costroc} becomes

\begin{equation}
\begin{split}
R = &\ P(t=1)\ \textit{FNR}\ C(0,1) \\
& + P(t=0)\ \textit{FPR}\ C(1,0)
\end{split}
\label{costroc1}
\end{equation}

Selecting the threshold which allows the best trade-off is done by drawing isocost lines as described in \cite{fawcett2006introduction}.
Using~\cref{costroc1}, it can be derived that two points $(1-\textit{FNR}_{1},\textit{FPR}_{1})$ and $(1-\textit{FNR}_{2},\textit{FPR}_{2})$, in~\cref{rocfig}, have the same performance if

\begin{equation}
\begin{split}
\frac{\textit{FNR}_{1}-\textit{FNR}_{2}}{\textit{FPR}_{2}-\textit{FPR}_{1}} & =\frac{C(1,0)\ P(t=0)}{C(0,1)\ P(t=1)} \\
& = m \text{ , with } m \in \mathbb{R}^{+\ast}
\end{split}
\label{ratio}
\end{equation}

In~\cref{rocfig}, the minimum isocost line $(d)$ is depicted for $m=\frac{2}{3}$.
Therefore, $m$ is tuned by the proportion of binary targets from training samples and by specific costs which can be presented in a cost matrix,

\begin{equation}
R =
\begin{pmatrix}
C(1,1) & C(0,1) \\
C(1,0) & C(0,0)
\end{pmatrix}
\end{equation}

Concretely, we set this cost matrix to

\begin{equation}
R =
\begin{pmatrix}
0 & 0.85 \\
0.15 & 0
\end{pmatrix}
\label{costmatchoice}
\end{equation}

\subsection{Reinforcement Learning}\label{reinfo}

To best deal with the outlined need for online learning, another feasible approach is \textit{reinforcement learning} that is tailored to learn without needing to be given a desired output.
This problem occurs, whenever the \ac{DPM} system decides to not decode the control channel of a \ac{TTI}.

A reinforcement learning system~\cite{sutton1998reinforcement}, in general, consists of an \emph{agent} that interacts with the \emph{environment}.
The agent performs an \emph{action} $a$, which will affect the environment leading to a new environment state $s$.
Subsequently, the agent will use this environment state $s$ in order to choose the next action.
As additional feedback, the agent receives a \textit{reward} $r$, which is either calculated from the observed state $s$ or is explicitly given by the environment.
The \textit{reward} indicates how desirable it is for the environment to be in this state.
The agent's goal is to choose actions that maximize the long term reward.

\begin{figure}
  \centering
  \resizebox{0.48\textwidth}{!}{\includegraphics{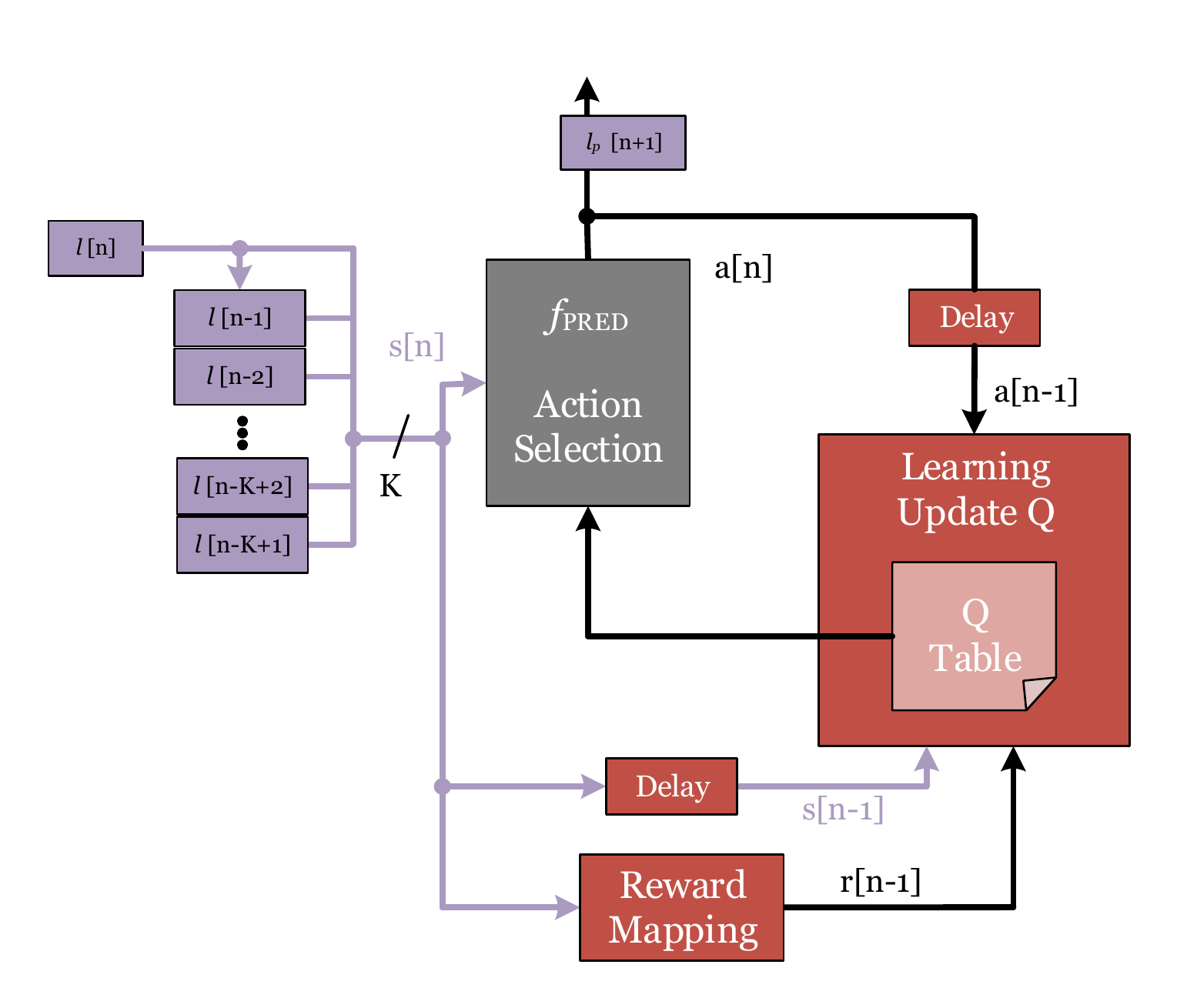}}
 
  \caption{\label{reinf}Reinforcement Learning-based prediction: For each possible combination of current state $\vec{s}[n]$ -- a sliding window of the last $K$ observed control information -- and action $a$ -- all possible scenarios according to~\cref{traces} -- the predictor keeps track of an estimated long-term reward -- the Q-value $Q(\vec{s}[n],a)$.
    At each time step, an action is chosen based on this Q-value (\cref{reinfpolicy}).
    To make sure that the estimated Q-value is accurate, the $Q(\vec{s}[n-1], \vec{a}[n-1])$ of the last state $\vec{s}[n-1]$ and last chosen action $\vec{a}[n-1]$ is updated (\cref{algos}) according to the reward $\vec{r}[n-1]$ received, which is deduced from $\vec{s}[n]$.
    To steer the agent towards desired behavior, problem-specific reward identification and assignment logic is crucial (\cref{rewards}) as well as appropriate Q-value initialization (\cref{qinit}).}
\end{figure}

The general approach works as follows:
For each possible pair $(s,a)$ of an observable state $s$ of the environment and each action $a$ that may be taken in this state, the agent stores an expected \emph{long-term reward} value $Q(s,a)$ for taking the action $a$ in state $s$.
Upon encountering a state $s$, the agent determines an appropriate action according to an \emph{action selection} algorithm, based on the stored $Q$-values.
After that, the agent calculates a reward $r$ (\emph{reward mapping}) based on the observed response from the environment and updates the estimated value of $Q(s,a)$ of the \textit{last} state and \emph{last} chosen action.
Because this update considers the $Q$-value of the \emph{next} state, even negative rewards, like retransmissions in our \ac{LTE} case, will propagate back through previous actions to the erroneous prediction.
In the following, we apply these notions and ideas of reinforcement learning to our time series-based task of online prediction of $\vec{l}[n+1]$ based on a known time series of $K$ previous TTI information tuples, see~\cref{reinf}.

Here, the state $\vec{s}[n]$ during TTI $n$ is given by the presence and absence of ULG and DLG in the current TTI control tuple information $\vec{l}[n]$ and the last $K-1$ previous tuples.
The action $\vec{a}[n]$ selected by the agent for the $n$th TTI corresponds to the prediction $\vec{l}_p[n+1]$ of the forthcoming TTI's control information according to the three scenarios distinguished in~\cref{powerman}.

\subsubsection{Action Selection}\label{reinfpolicy}
In our case of predictive LTE grant prediction, the observed training data may itself be affected by the agent's chosen previous actions.
Thus, an area of tension naturally arises: the balance of \emph{exploration} and \emph{exploitation}.

Exploration, on the one hand, refers to gathering information of the state and action space.
This is generally achieved by choosing actions in states that have not yet, or rather seldom, been visited.
Because the $Q(s,a)$-values cannot be initialized perfectly without prior learning, this means that exploration hazards the consequences of taking initially presumably suboptimal actions.
The trade-off is the potential to find better Q-values, or at the very least broadening the agent's knowledge for the future.
Exploitation, on the other hand, is the simple act of taking the -- currently estimated -- best action $\vec{a}^\mathrm{best}$, i.e. the action that maximizes the expected long-term reward:

\[ \vec{a}^\mathrm{best} = \underset{a}{\argmax}\;Q(s,a) \]

Obviously, a suitable balance between exploitation and exploration is key to success when applying reinforcement learning.
A handicap in our context of LTE grant  prediction is that important information for taking the right decision on \ac{DPM} is not directly observable by a single UE, like radio condition, number of other UEs in a cell, and the scheduling strategy of the eNodeB.
Indeed, in dynamic environments, these unknowns even undergo a constant flux.
We therefore argue for a strategy for action selection that permanently has the capacity to explore and refine.

For action selection, an $\epsilon$-Greedy strategy \cite{watkins1992q} is proposed, which is a function that chooses at the $n$th TTI the action with the highest estimated long-term reward value $\vec{a}^\mathrm{best}$ with a chance of $1-\epsilon$, and a random action with a chance $\epsilon$.

\begin{equation}
\vec{a}[n] = \begin{cases}
\textrm{random action}   \quad\text{with probability }\epsilon \\
\vec{a}^\mathrm{best} \quad\quad\quad\quad\quad\,\,\text{with probability }1-\epsilon
\end{cases}
\end{equation}
Thereby, the parameter $\epsilon$ can be set to a constant, e.g., $\epsilon= 10\,\%$ in our experiments.
Alternatively, the value could even be adapted online to explore different behavior (if the trace characteristics change) or to exploit more often (if the trace characteristics are stable).

\subsubsection{Dedicated Learning Phase}\label{ddlp}
As we will show in~\cref{experiments}, the time an agent might need to achieve acceptable \acp{FNR} may be significant.
Because the UE is operated in an LTE cell that potentially penalizes UEs that exhibit a disproportionately large rate of missed messages, we propose to start up with an initial dedicated learning phase.
During this phase, the actions proposed by the agent are not forwarded to the \ac{DPM}.
Instead, existing reactive LTE DPM policies are used.
However, the rewards are calculated as if the agent decision were used.
For example, if scenario $z_3$ (\emph{No DLG $\land$ No ULG} according to~\cref{traces}) is predicted, we will not turn off the RF\_RX and PHY\_RX components immediately.
Rather, if a grant should be received within the TTI, a negative reward (for theoretically missing the information) is issued.

During this phase, a moving average of the \ac{FNR} is calculated.
Only after reaching a minimal error threshold, in our case of $\epsilon^\mathrm{min\_err} = 40\%$, the system starts powering the system components down according to the actions as suggested by the agent.
Furthermore, we introduce a maximal error threshold $\epsilon^\mathrm{err\_max} = 45\,\%$, that upon reaching will trigger a new dedicated learning phase, to account for recognizing changes and transients in trace characteristics.
Of course, this introduces a certain amount of time where no energy can be saved.
This is offset by a guaranteed worst-case impact, because no grants will be missed.

\subsubsection{LTE-Specific Reward Mapping Algorithm}\label{rewards}
After choosing an action, the agent maps the response of the environment to a reward $\vec{r}[n-1]$.
This is realized by a mapping function that checks reality $\vec{l}[n]$ against the last prediction $\vec{l}_p[n]$ for desirable and undesirable attributes.

Desirable are all actions that minimize energy consumption, i.e., that favor turning components off as early as possible.
Undesirable are in descending order of severity: (i) turning components off too early (false negative error) and (ii) turning components off too late, or not at all (different degrees of false positive errors).

The proposed reward assignment is shown in~\cref{rewardstable} ordered by the prediction $\vec{l}_p[n]$, i.e., the last action $\vec{a}[n-1]$ performed by the agent.
For the cases $\vec{l}_p[n] \in \{(\true,\false), (\true,\true), (\false,\true)\}$, the modem will at least receive and decode the control channel, which allows the evaluation of whether data has been lost or an opportunity to save energy was missed.
The prediction $\vec{l}_p[n] = (\false,\false)$ is special, as it leads to the TTI information $\vec{l}[n]$ not being received, meaning no direct evaluation of prediction accuracy is possible.

\begin{table}[bth]
  \centering
  \begin{tabular}{cccc}
    \toprule
       Prediction                                     & Reality                        & Reward       & Description     \\
       $\vec{l}_p[n]$                                 & $\vec{l}[n]$                   &$\vec{r}[n-1]$&                 \\
    \midrule
      \multirow{3}{*}{$(\true,\false)$}               & $(\true,\false)$               & 2            & energy saved         \\ 
                                                      & $(\true,\true) (\false,\true)$ & -5          & false negative  \\ 
                                                      & $(\false,\false)$              & 0           & false positive  \\ 
    \midrule
      \multirow{3}{*}{$(\true,\true), (\false,\true)$}& $(\true,\false)$               & 0            & false positive  \\ 
                                                      & $(\true,\true) (\false,\true)$ & 2            & energy saved         \\ 
                                                      & $(\false,\false)$              & 0           & false positive  \\ 
    \midrule
       \multirow{4}{*}{$(\false,\false)$}             & $(\true,\false)$               &  -5          & false negative  \\
                                                      & $(\true,\true) (\false,\true)$ & -5          & false negative  \\
                                                      & $(\false,\false)$              &  0           & energy saved         \\
                                                      & $(\unknown,\unknown)$          &  0           & assumed correct \\ 
    \bottomrule
  \end{tabular}
  \vspace{1mm}
  \caption{\label{rewardstable}Deriving the reward $\vec{r}[n-1]$ for the last action $\vec{a}[n-1] \equiv \vec{l}_p[n]$ and the really observed TTI information $\vec{l}[n]$.
    The reward value is in descending order: (i) correct predictions, (ii) false positives, and (iii) false negatives.
    A special case is $\vec{l}_p[n] = (\false,\false)$ as a direct assessment of prediction quality can only be performed during the dedicated learning phase.
    During the exploitation phase, the input $\vec{l}[n]$ may sometimes not be observable $(\unknown,\unknown)$, i.e., our predictive DPM decided to completely turn of the modem.
    In this case, we assume the prediction to be correct and assign a positive reward.
    However, if turning the modem of turned out to be wrong, i.e, data had to be re-transmitted by the \ac{eNodeB}, the misprediction will be penalized later through the discussed additional mechanisms.}
\end{table}

However, to make sure a wrong prediction of $\vec{l}_p[n]=(\false,\false)$ is discouraged, we further introduce three \emph{additional mechanisms}:
(i) If the \ac{UE} has received a DLG with $ndi \equiv \false$, a negative reward $r_\mathrm{ndi} = -5 $ is awarded.
As explained in~\cref{ltebackground}, this indicates that a prior DLG from the \ac{eNodeB} was missed, indicating a wrong prediction of $(\false,\false)$ some time in the past.
(ii) The agent may not predict $(\false,\false)$ more than $K = 3$ times in a row.
This ensures that the agent does not permanently turn off the modem (obtaining the reward $r_\mathrm{off}$) while evading the first mechanism (by making communication impossible).
(iii) Assigning the negative reward $r_\mathrm{bsr} = -5$ if no ULG was received for a BSR (see~\cref{sec::uplink}) sent within the last $10$ \acp{TTI}.
This situation may indicate that control information -- the missing ULG -- was lost some time in the past.

\subsubsection{Learning}\label{algos}
Based on the calculated reward $\vec{r}[n-1]$, the $Q$-values are updated.

For this, we propose to employ SARSA-$\lambda$ (see~\cite{rummery1994line}).

\begin{equation}\label{sarsa}
\begin{aligned}
  \scalebox{.95}{$Q(\vec{s}[n-1], \vec{a}[n-1])$} &= \scalebox{.95}{$(1-\alpha)\cdot Q(\vec{s}[n-1], \vec{a}[n-1])$}\\
                                                  &+ \scalebox{.95}{$\alpha\cdot (Q(\vec{s}[n], \vec{a}[n])\cdot \gamma + \vec{r}[n-1])$}
\end{aligned}
\end{equation}
It updates the respective Q-values after having observed the immediately experienced reward $\vec{r}[n-1]$ resulting from the last action $\vec{a}[n-1]$ with this reward and the estimated long-term reward $Q(\vec{s}[n], \vec{a}[n])$ (of the current state) to a degree determined by $\alpha$.
We chose SARSA-$\lambda$ over other simple algorithms like Q-Learning, as SARSA-$\lambda$ generally penalizes actions leading to bad rewards stronger.

\subsubsection{Initialization of Q-Values}\label{qinit}
The initialization of the $Q$-values can quite significantly affect the initial prediction quality.
As explained before, the reception of the control channel information allows for the most accurate reward assignment, because the predictor can deduce the perfect action.
Considering that, a Q-value initialization favoring $\vec{l}_p[n+1] = (\false, \false)$ is highly discouraged.
Additionally, as outlined in~\cref{class}, false negative errors may have a significant impact on transmission quality and speed of the whole cell.
Therefore, a setup minimizing the FNR seems appropriate.
To realize this, we propose an initialization in the following order:
\begin{equation*}
  \scalebox{.95}{$\forall s \in S: Q(s, (\true,\true)) = Q(s, (\false,\true)) \ge Q(s, (\true,\false)) \ge Q(s, (\false,\false))$}
\end{equation*}
The discrepancy in value must be small enough to allow for a fast adaption to new experiences.

\section{Evaluation}\label{experiments}
This section compares both presented DPM approaches, regarding both functional and non-functional properties.
First, in~\cref{subsec:framework}, we give a short introduction to our simulation-based evaluation framework.
Next, the application of video streaming is chosen as introduced in~\cref{subsec::funcmodel}.
For trace characterization, we introduce suitable metrics in~\cref{subsec:metrics}.
Subsequently, we perform a complexity analysis (\cref{subsec:complexity}) based on \acp{FLOP} used for training and prediction. 
Finally, in~\cref{subsec:results}, an in-depth evaluation and comparison of both approaches in terms of accuracy, learning rates, and energy savings is presented based on the previously introduced concepts.

\subsection{Simulation Framework}\label{subsec:framework}

Designing a predictive \ac{DPM} system poses certain constraints on the nature of the realization.
Apparently, such prediction techniques must be of low computation complexity, but also lead to net overall energy savings in order to be economically of interest.
Additionally, in contrast to reactive power management systems, employing a prediction step inherently introduces a certain degree of uncertainty, as explained in \cref{class}.

For the evaluation of potential energy savings and scope of mispredictions of our proposed predictive \ac{DPM}, we need an evaluation that reflects a UE in a real cell environment with a sophisticated energy modeling. 
For a realistic and parameterizable model of the \ac{LTE} cell environment, we employ the ns-3 simulator~\cite{Riley2010}.
In order to quantify the energy consumption (\cref{formal}) of the hardware model (\cref{sec:arch}), a SystemC-based simulator~\cite{rgtwxh_2014-maestro} is used.
Both simulators exchange relevant information on a subframe basis through a cosimulation interface.

\subsection{Functional Application Model}\label{subsec::funcmodel}
As both predictive \ac{DPM} techniques are designed to be trained in a live network cell, there are two main characteristics that need to be considered.
The first characteristic is the length of stable trace behavior.
Obviously, if the grant patterns change too quickly, a predictor will always be stuck in the learning phase, without ever being able to get into the exploitation phase.
Obviously, no energy can be saved in such circumstances.
It will be shown that the required minimum length of this stable interval depends on the patterns in the trace themselves and the employed approach.

The second characteristic is the grant density, i.e., the proportion of \acp{TTI} that contain either an ULG, DLG, or both, which is explained in~\cref{subsec:metrics}.

In the following experiments, we carefully investigate traces and prediction accuracy for highly different manifestations of each characteristic.
Based on a parameterizable video streaming application, we create a variety of different scenarios with varying transmission length (seconds of short video playback or several minutes long videos) and varying resolutions corresponding to different data rates.

Our modeling realizes the proposed video stream algorithm from \cite{ameigeiras2012analysis}, which is given in \cref{youtubeapplication}.
\begin{algorithm}
\SetAlgoLined\DontPrintSemicolon

\SetKwInOut{Input}{input}\SetKwInOut{Output}{output}
\SetKwFunction{proc}{Initialization}
\SetKwFunction{fillproc}{Filling}
\SetKwProg{myproc}{Procedure}{}{}
\Input{$S, R_\text{encoding}$}
\myproc{\proc{}}{
$R_{sending} \gets R_{encoding}\cdot 1.25$\;
$S_B \gets 5\cdot R_{encoding}$\;
$S \gets S -S_B$\;
\BlankLine
\emph{socket\_open($2\cdot 10^6$)}\;
\While{$S_B \ge 0$}{
\emph{socket\_send(P)\;}
}
\emph{Filling()\;}
}
\BlankLine
\myproc{\fillproc{}}{
$t_\text{steady} \gets \frac{P}{R_\text{sending}}$\;
\While{$ S \ge 0$}{
\emph{sleep($t_\text{steady}$)\;}
\emph{socket\_send(P)\;}
$S \gets S - P\;$
}
}

\caption{Server application model from \cite{ameigeiras2012analysis}.}\label{youtubeapplication}
\end{algorithm}
Here, a video transmission of a video of size $S$ bytes consists of a first burst phase (\emph{Initialization procedure}), followed by smaller periodic transmissions during the \emph{Filling procedure}.
During the initialization procedure, in total $S_B = 5\cdot R_\text{encoding}$ bytes, corresponding to approximately 40s of encoded video length, are transmitted at maximum speed.
During the filling procedure, a packet of size $ P = 64$ KB is transmitted every $t_\text{steady} = \frac{P}{R_{sending}}\cdot 10^3$ ms time duration, with a sending rate of $R_{sending} = R_{encoding}\cdot 1.25$.
For realistic values of the encoding rate, we use the values as reported in \cite{ameigeiras2012analysis}, with $R_\text{encoding} \in [200\frac{\text{KB}}{\text{s}}, 3320\frac{\text{KB}}{\text{s}}]$.

\subsection{Trace Characterization}\label{subsec:metrics}
The evaluation of the prediction techniques is performed for traces of lenghts up to $10^6$ TTIs in the following, varying in the \emph{grant density} of both ULGs and DLGs.
We define the grant densities $D_\mathrm{DLG}$ and $D_\mathrm{ULG}$ for a time series of length $K$ as the proportion of \acp{TTI} containing the respective kind of grant.

\begin{equation}
D_G^K = \dfrac{\#\text{ of grants of type G}}{K}, \;\;\;\;G \in \{ \mathrm{DLG}, \mathrm{ULG}\}
\end{equation}

These densities define upper bounds in terms of achievable energy savings of each of the predictive DPM approaches.
To exemplify, consider a trace with a $D_\mathrm{DLG}= 100\,\%$, i.e., a DLG in every \ac{TTI}.
Obviously, not even an ideal, perfect predictor could reduce the energy consumption compared to a reactive \ac{DPM}, because each channel of every \ac{TTI} contains information and has to be decoded.
For a trace with $D_\mathrm{DLG} = 50\,\%$, on the other hand, the proposed predictive \ac{DPM} approaches could optimize the behavior for the remaining $50\,\%$ compared to a naive approach.

\subsection{Complexity Analysis}\label{subsec:complexity}

\begin{figure}[bth]
  \centering  
  \includegraphics[width = .39\textwidth]{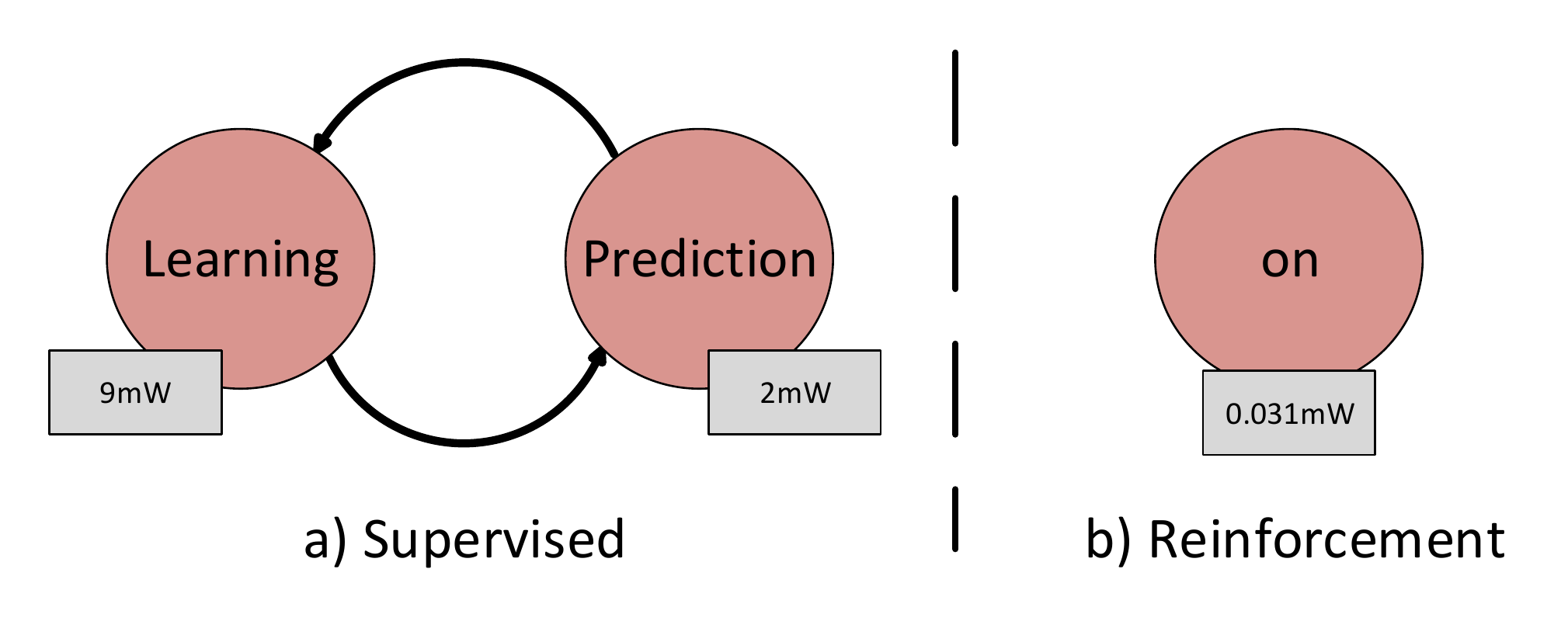}
  \caption{\label{psmpredictor}Power State Machines of both the a) supervised and b) reinforcement predictors.
    For the supervised predictor, the computational complexity, and thus the power consumption differs between a higher power state -- during the training phase -- and a lower power state -- during the exploitation phase.
    The reinforcement predictor is always in a learning phase, and thus has a constant power consumption.
    The power consumption is approximated from the calculation of FLOPs per TTI, assuming their implementation in software on a DSP for modems~\cite{anjum2011state}.}
\end{figure}

As target platform for testing the presented prediction algorithms, we assume given an LTE base band \ac{DSP} from literature, as discussed in~\cite{anjum2011state, daly2017through}, that runs at a clock frequency of $f = 300~\text{MHz} $ with a power consumption of $P_f = 1\,\text{mW/MHz} $ (including the power consumption of its memory).
This \ac{DSP} introduces another component (with a power state machine, see~\cref{psmpredictor}) to the architecture model introduced in \cref{powerman}.
To estimate the inherent energy overhead $E_Q$ of the presented algorithms per \ac{TTI}, we propose to employ a \ac{FLOP}-based approach.
Here, we translate one iteration (that is performed in each case per \ac{TTI}) of the presented algorithms first to the number of required \acp{FLOP} $c_Q$ (according to~\cref{tab:opToFlop}, from \cite{wu2012performance}) and second to power consumption.
We assume that both the Q-table as well as the operations and weights describing the neural net fit on the on-chip memory of the \ac{DSP}.

\begin{table}[tbh]
  \centering
 
  \begin{tabular}{cc} 	  
 
    \toprule
      Operation      & Complexity $[\text{\ac{FLOP}}]$  \\
    \midrule
      Addition       & 1 \\
      Subtraction    & 1 \\
      Comparison     & 1 \\
      Multiplication & 2 \\
      Division       & 4 \\
      Exponential    & 8 \\
    \bottomrule
  \end{tabular}
  
  \vspace{1mm}
  \caption{\label{tab:opToFlop}Translation of arithmetic operations to \acp{FLOP} count.}
  \vspace{1mm}
\end{table}

Based on the assumption of one FLOP per DSP cycle, $E_Q$ per \ac{TTI} can be calculated according to:
\begin{equation}
  E_Q = \frac{c_Q}{f} \cdot P_f \cdot f
\end{equation}

If we assume that the predictor is run for the duration of the whole TTI, i.e., $1$\,ms, this yields the power consumption $P_Q = \frac{E_Q}{1\,\textrm{ms}}$ as modeled by a power state machine for the predictor (see~\cref{psmpredictor}).

For the supervised predictor, we distinguish between two power states, corresponding to (i) the learning phase and (ii) the exploitation phase, as both phases differ in algorithm and thus number of computations.
For the FFNN introduced in~\cref{superv}, we obtain $7,018$ \acp{FLOP} per TTI for learning, and $2,014$ \acp{FLOP} per TTI during the exploitation phase.
Because we perform a prediction alongside each learning step to evaluate the prediction accuracy, one complete iteration during learning requires a total of $7,018+2,014 = 9,032$ \acp{FLOP}.
With the introduced \ac{DSP} model, this translates to an energy consumption per \ac{TTI} of $E_Q^L = 9\,\mu J$ for learning and $E_Q^P = 2\,\mu J$ for exploitation, respectively.
Thus, the two power states in~\cref{psmpredictor} corresponding to a power consumption of $P_Q^L = 9\,\text{mW}$ and $P_Q^P = 2\,\text{mW}$, respectively.

The reinforcement predictor, always updating its Q-values according to~\cref{sarsa}, respectively, is always in the same power state.
One step for Q-learning requires $19$ \acp{FLOP}, while one step of sarsa-$\lambda$ requires $25$ \acp{FLOP}.
Factoring in the reward mapping function requiring $12$ comparisons, translating to $12$ \acp{FLOP}.
In sum sarsa-$\lambda$ requires  $37$ \acp{FLOP} per \ac{TTI} translating to an energy consumption of $E_Q^{S} = 0.037\,\mu \text{J}$ resulting in power states with a power consumption of $P_Q^S = 0.037\,\text{mW}$.

\subsection{Results}\label{subsec:results}
This section discusses experimental results that were obtained by both prediction approaches in the presented simulation environment.
In \cref{sec::predAccuracy}, we investigate learning time and prediction accuracy in terms of \ac{FNR} of both approaches for different scenarios.
\cref{sec::energyConsump} then investigates how this affects the modem energy consumption.

We present the results for three representative traces $\vec{l}_i \in \{\vec{l}_\text{min}, \vec{l}_\text{avg} ,\vec{l}_\text{max}\}$ defined by their respective encoding rate $R_\text{encoding}^{\vec{l}_i}$, covering a large spectrum of different grant densities $D_G$:
\begin{enumerate}
  \setlength\itemsep{1em}
	\item[(i)] $R_\text{encoding}^{\vec{l}_\text{min}} = 200 \frac{\text{KB}}{\text{s}}$, with $D_\text{DLG} = 0.299507$
	\item[(ii)] $R_\text{encoding}^{\vec{l}_\text{avg}} = 3320\frac{\text{KB}}{\text{s}}$, with $D_\text{DLG} = 0.652489$
	\item[(iii)] $R_\text{encoding}^{\vec{l}_\text{max}} = \dfrac{3320 + 200}{2}\frac{\text{KB}}{\text{s}} = 1760\frac{\text{KB}}{\text{s}}$, with $D_\text{DLG} = 0.827384$
\end{enumerate}

\subsubsection{Prediction Accuracy}\label{sec::predAccuracy}
First, we evaluate the prediction accuracy in terms of FNR, giving an indication of the maximum negative impact on the whole cell.
\begin{figure*}[!t]
\centering
\begin{subfigure}[t]{.45\textwidth}
\scalebox{.8}
{\begin{tikzpicture} \begin{axis}[
      xlabel={Interval[s]},
      ylabel=$FNR$]
\addplot[color=blue,line width = 1pt] plot coordinates {
(1,0.45)
(2,0.21)
(3,0.12)
(4,0.18)
(5,0.03)
(10,0.11)
(15,0.13)
(20,0.14)
(25,0.1)
(30,0.13)
(35,0.11)
(40,0.09)
(45,0.12)
(50,0.12)
(55,0.12)
(60,0.12)
(65,0.11)
(70,0.11)
(75,0.11)
(80,0.1)
(85,0.12)
(90,0.1)
(95,0.12)
};
\addplot[color=blue,dashed, line width = 1pt] plot coordinates { 
(1,0.42)
(2,0.14)
(3,0.05)
(4,0.02)
(5,0.02)
(10,0.08)
(15,0.13)
(20,0.1)
(25,0.1)
(30,0.12)
(35,0.08)
(40,0.1)
(45,0.08)
(50,0.12)
(55,0.07)
(60,0.12)
(65,0.1)
(70,0.09)
(75,0.11)
(80,0.08)
(85,0.1)
(90,0.08)
(95,0.1)
};

\addplot[color=green, dashed,line width = 1pt] plot coordinates {
(1,0.42)
(2,0.15)
(3,0.06)
(4,0.02)
(5,0)
(10,0.04)
(15,0.07)
(20,0.03)
(25,0.02)
(30,0.04)
(35,0.03)
(40,0.03)
(45,0.03)
(50,0.03)
(55,0.03)
(60,0.03)
(65,0.03)
(70,0.03)
(75,0.03)
(80,0.03)
(85,0.03)
(90,0.03)
(95,0.03)
};
\addplot[color=green,line width = 1pt] plot coordinates {
(1,0.46)
(2,0.31)
(3,0.12)
(4,0.16)
(5,0.02)
(10,0.12)
(15,0.11)
(20,0.11)
(25,0.13)
(30,0.12)
(35,0.11)
(40,0.11)
(45,0.11)
(50,0.1)
(55,0.11)
(60,0.1)
(65,0.1)
(70,0.1)
(75,0.1)
(80,0.09)
(85,0.08)
(90,0.08)
(95,0.09)
};

\addplot[color=violet,dashed,line width = 1pt] plot coordinates {
(1,0.42)
(2,0.16)
(3,0.09)
(4,0.03)
(5,0)
(10,0.02)
(15,0.02)
(20,0.02)
(25,0.02)
(30,0.02)
(35,0.02)
(40,0.02)
(45,0.02)
(50,0.02)
(55,0.02)
(60,0.02)
(65,0.02)
(70,0.02)
(75,0.02)
(80,0.02)
(85,0.02)
(90,0.02)
(95,0.02)
};
\addplot[color=violet,line width = 1pt] plot coordinates {
(1,0.45)
(2,0.3)
(3,0.14)
(4,0.18)
(5,0.02)
(10,0.12)
(15,0.1)
(20,0.1)
(25,0.09)
(30,0.1)
(35,0.1)
(40,0.1)
(45,0.09)
(50,0.09)
(55,0.09)
(60,0.08)
(65,0.08)
(70,0.08)
(75,0.08)
(80,0.09)
(85,0.08)
(90,0.08)
(95,0.08)
};

\end{axis};

\end{tikzpicture}}
  \caption{\label{fig:fnrdebug} No rescheduling of missed transmissions.}
    \end{subfigure}
    \begin{subfigure}[t]{.45\textwidth}
    \scalebox{.8}
      {\begin{tikzpicture} \begin{axis}[
      xlabel={Interval[s]},
      ylabel=$FNR$]
\addplot[color=violet,line width = 1pt] plot coordinates { 
(1,0.46)
(2,0.3)
(3,0.13)
(4,0.18)
(5,0.02)
(10,0.24)
(15,0.23)
(20,0.2)
(25,0.2)
(30,0.21)
(35,0.19)
(40,0.14)
(45,0.13)
(50,0.16)
(55,0.14)
(60,0.15)
(65,0.13)
(70,0.14)
(75,0.16)
(80,0.15)
(85,0.14)
(90,0.15)
(95,0.14)
};
\addplot[color=violet,dashed,line width = 1pt] plot coordinates {
(1,0.43)
(2,0.18)
(3,0.08)
(4,0.02)
(5,0)
(10,0.34)
(15,0.23)
(20,0.17)
(25,0.04)
(30,0.22)
(35,0.22)
(40,0.17)
(45,0.13)
(50,0.1)
(55,0.09)
(60,0.18)
(65,0.19)
(70,0.1)
(75,0.1)
(80,0.13)
(85,0.04)
(90,0)
(95,0)
};

\addplot[color=blue,line width = 1pt] plot coordinates {
(1,0.45)
(2,0.21)
(3,0.1)
(4,0.18)
(5,0.03)
(10,0.23)
(15,0.2)
(20,0.24)
(25,0.18)
(30,0.21)
(35,0.19)
(40,0.14)
(45,0.14)
(50,0.15)
(55,0.15)
(60,0.13)
(65,0.16)
(70,0.16)
(75,0.14)
(80,0.14)
(85,0.15)
(90,0.12)
(95,0.15)
};
\addplot[color=blue,dashed,line width = 1pt] plot coordinates {
(1,0.42)
(2,0.12)
(3,0.04)
(4,0.02)
(5,0.01)
(10,0.25)
(15,0)
(20,0.21)
(25,0.09)
(30,0.33)
(35,0.23)
(40,0.23)
(45,0.17)
(50,0.12)
(55,0.08)
(60,0.14)
(65,0.19)
(70,0.15)
(75,0.24)
(80,0.28)
(85,0.07)
(90,0)
(95,0)
};
\addplot[color=green,line width = 1pt] plot coordinates {
(1,0.45)
(2,0.18)
(3,0.11)
(4,0.17)
(5,0.02)
(10,0.25)
(15,0.28)
(20,0.23)
(25,0.19)
(30,0.16)
(35,0.16)
(40,0.16)
(45,0.17)
(50,0.16)
(55,0.14)
(60,0.18)
(65,0.13)
(70,0.16)
(75,0.14)
(80,0.15)
(85,0.15)
(90,0.14)
(95,0.14)
};
\addplot[color=green,dashed,line width = 1pt] plot coordinates {
(1,0.42)
(2,0.13)
(3,0.04)
(4,0.02)
(5,0)
(10,0.28)
(15,0.41)
(20,0.41)
(25,0.41)
(30,0.29)
(35,0.24)
(40,0.16)
(45,0.26)
(50,0.2)
(55,0.15)
(60,0.11)
(65,0.13)
(70,0.21)
(75,0.2)
(80,0.16)
(85,0.26)
(90,0.17)
(95,0.19)
};
\end{axis};
\end{tikzpicture}}
      \caption{\label{fig:fnrempty} Rescheduling of missed transmissions.}
        \end{subfigure}
         \caption{\label{fig:fnrreinffig} Depicted is the FNR calculated as moving average over a window of the last 3,000\,ms for 3 representative traces.
            The reinforcement FNR is shown as a smooth and the supervised FNR as a dashed line. The scenarios are distinguished by color: blue corresponds to $\vec{l}_\text{min}$, green to $\vec{l}_\text{avg}$, and violet to $\vec{l}_\text{max}$.}
\end{figure*}
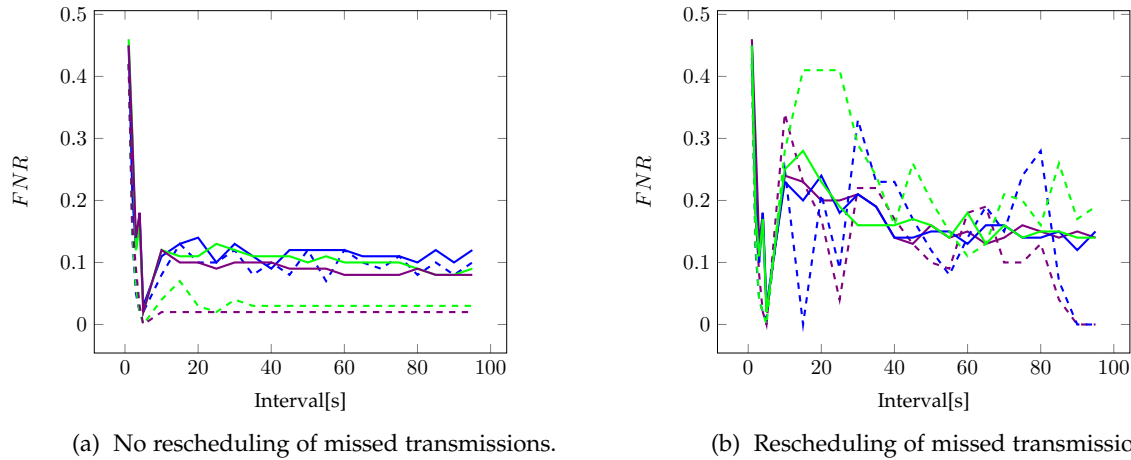
Reflecting our time-series problem, we evaluate these values absolutely, as well as their shift over time.
\cref{fig:fnrreinffig} shows the calculation of the FNR as a moving average over intervals of $3,000$ \acp{TTI}.
To first showcase the prediction accuracy, this evaluation is performed while (a) disregarding changed trace characteristics due to missed transmissions and (b) rescheduling missed traffic.

For the first $5,000$ \acp{TTI}, the average FNR is high, strengthening our argument for a dedicated learning phase.
Afterwards, for stable traces (\cref{fig:fnrdebug}), both prediction approaches quickly converge to desirably low \acp{FNR}.
Here, the supervised approach, cognizant of all prediction errors, outperforms the reinforcement predictor especially for both $\vec{l}_\text{min}$ and $\vec{l}_\text{max}$, achieving \acp{FNR} of lower than $1\%$.
Due to the $\epsilon$-Greedy strategy used, and the only grant presence-based prediction, the reinforcement predictor exhibits a higher \ac{FNR}.
Both approaches achieve stable prediction rates of lower than $15\%$.
For traces that are subject to changes due to missing traffic (\cref{fig:fnrempty}), the first characteristic we observe is the added difficulty for prediction, as the \ac{FNR} of both predictors, for all scenarios are significantly higher (but still lower than $\epsilon^\text{min\_err}$).
While the initial convergence remains fast, upon exiting the dedicated learning phase, re-transmissions occur, especially for $\vec{l}_\text{max}$, with the highest density $D_\text{DLG}$.
While this peak in errors occurs for both approaches and all scenarios, a far more severe oscillation for the supervised predictor is observable due to no mechanism to cope with the information contained in re-transmitted grants.
The reinforcement predictor, however, quickly incorporates this information of re-transmissions into subsequent predictions as outlined in ~\cref{rewards}, show-casing the online learning capabilities.
Despite their differences, both learning approaches achieve the desired task of learning grant prediction patterns. 
For all shown traces, the exploitation phase is reached, allowing for energy to be saved by following the predictions, as shown in~\cref{sec::energyConsump}.

\subsubsection{Energy Consumption}\label{sec::energyConsump}
\begin{figure}[t]
  \centering
  \scalebox{.8}{\begin{tikzpicture} 
 \begin{axis}[ 
      ytick={.85,.9,.95,1},
      xlabel={Interval$[s]$},
      ylabel={Energy normalized to $E^\text{DPM}$}, legend pos=north east]

\addplot[dashed,color=green,line width = 1pt] plot coordinates { 
(1,1.03)
(2,1.03)
(3,1.02)
(4,1.02)
(5,1.02)
(6,1.01)
(7,1.01)
(8,0.99)
(9,0.98)
(10,0.97)
(11,0.96)
(12,0.95)
(13,0.95)
(14,0.94)
(15,0.94)
(16,0.93)
(17,0.93)
(18,0.93)
(19,0.93)
(20,0.93)
(21,0.93)
(22,0.93)
(23,0.94)
(24,0.94)
(25,0.94)
(26,0.94)
(27,0.93)
(28,0.93)
(29,0.93)
(30,0.93)
(31,0.93)
(32,0.93)
(33,0.94)
(34,0.94)
(35,0.94)
(36,0.94)
(37,0.94)
(38,0.94)
(39,0.94)
(40,0.94)
(41,0.94)
(42,0.94)
(43,0.94)
(44,0.94)
(45,0.94)
(46,0.94)
(47,0.94)
(48,0.94)
(49,0.94)
(50,0.94)
(51,0.94)
(52,0.94)
(53,0.94)
(54,0.94)
(55,0.94)
(56,0.94)
(57,0.94)
(58,0.94)
(59,0.94)
(60,0.94)
(61,0.94)
(62,0.94)
(63,0.94)
(64,0.94)
(65,0.94)
(66,0.94)
(67,0.94)
(68,0.94)
(69,0.94)
(70,0.94)
(71,0.94)
(72,0.94)
(73,0.94)
(74,0.94)
(75,0.94)
(76,0.94)
(77,0.94)
(78,0.94)
(79,0.94)
(80,0.94)
(81,0.94)
(82,0.94)
(83,0.94)
(84,0.94)
(85,0.94)
(86,0.94)
(87,0.94)
(88,0.94)
(89,0.94)
(90,0.94)
(91,0.94)
(92,0.94)
(93,0.94)
(94,0.94)
(95,0.94)
(96,0.94)
(97,0.94)
(98,0.94)
(99,0.94)
};
\addplot[dashed,color=violet,line width = 1pt] plot coordinates { 
(1,1.03)
(2,1.03)
(3,1.02)
(4,1.02)
(5,1.02)
(6,1.01)
(7,1.01)
(8,1)
(9,1)
(10,1)
(11,0.99)
(12,1)
(13,0.99)
(14,0.99)
(15,0.99)
(16,0.99)
(17,0.99)
(18,0.99)
(19,0.99)
(20,0.99)
(21,0.99)
(22,0.99)
(23,0.99)
(24,0.99)
(25,0.99)
(26,0.99)
(27,0.98)
(28,0.99)
(29,0.98)
(30,0.98)
(31,0.98)
(32,0.98)
(33,0.98)
(34,0.98)
(35,0.98)
(36,0.98)
(37,0.98)
(38,0.98)
(39,0.98)
(40,0.98)
(41,0.98)
(42,0.98)
(43,0.98)
(44,0.98)
(45,0.98)
(46,0.98)
(47,0.98)
(48,0.98)
(49,0.98)
(50,0.98)
(51,0.98)
(52,0.98)
(53,0.98)
(54,0.98)
(55,0.98)
(56,0.98)
(57,0.98)
(58,0.98)
(59,0.98)
(60,0.98)
(61,0.98)
(62,0.98)
(63,0.98)
(64,0.98)
(65,0.98)
(66,0.98)
(67,0.98)
(68,0.98)
(69,0.98)
(70,0.98)
(71,0.98)
(72,0.98)
(73,0.98)
(74,0.98)
(75,0.98)
(76,0.98)
(77,0.98)
(78,0.98)
(79,0.98)
(80,0.98)
(81,0.98)
(82,0.98)
(83,0.98)
(84,0.98)
(85,0.98)
(86,0.98)
(87,0.98)
(88,0.98)
(89,0.98)
(90,0.98)
(91,0.98)
(92,0.98)
(93,0.98)
(94,0.98)
(95,0.98)
(96,0.98)
(97,0.98)
(98,0.98)
(99,0.98)
};
\addplot[dashed,color=blue,line width = 1pt] plot coordinates { 
(1,1.03)
(2,1.03)
(3,1.02)
(4,1.02)
(5,1.02)
(6,1)
(7,0.98)
(8,0.96)
(9,0.94)
(10,0.93)
(11,0.92)
(12,0.91)
(13,0.9)
(14,0.89)
(15,0.89)
(16,0.89)
(17,0.88)
(18,0.88)
(19,0.88)
(20,0.87)
(21,0.87)
(22,0.87)
(23,0.86)
(24,0.86)
(25,0.86)
(26,0.86)
(27,0.86)
(28,0.86)
(29,0.85)
(30,0.85)
(31,0.85)
(32,0.85)
(33,0.85)
(34,0.85)
(35,0.85)
(36,0.85)
(37,0.85)
(38,0.84)
(39,0.84)
(40,0.84)
(41,0.84)
(42,0.84)
(43,0.84)
(44,0.84)
(45,0.84)
(46,0.84)
(47,0.84)
(48,0.84)
(49,0.84)
(50,0.84)
(51,0.84)
(52,0.84)
(53,0.84)
(54,0.84)
(55,0.84)
(56,0.83)
(57,0.83)
(58,0.83)
(59,0.83)
(60,0.83)
(61,0.83)
(62,0.83)
(63,0.83)
(64,0.83)
(65,0.83)
(66,0.83)
(67,0.83)
(68,0.83)
(69,0.83)
(70,0.83)
(71,0.83)
(72,0.83)
(73,0.83)
(74,0.83)
(75,0.83)
(76,0.83)
(77,0.83)
(78,0.83)
(79,0.83)
(80,0.83)
(81,0.83)
(82,0.83)
(83,0.83)
(84,0.83)
(85,0.83)
(86,0.83)
(87,0.83)
(88,0.83)
(89,0.83)
(90,0.83)
(91,0.83)
(92,0.83)
(93,0.83)
(94,0.83)
(95,0.83)
(96,0.83)
(97,0.83)
(98,0.83)
(99,0.83)
};
\addplot[color=green,line width = 1pt] plot coordinates { 
(1,1)
(2,1)
(3,1)
(4,1)
(5,1)
(6,0.99)
(7,0.98)
(8,0.96)
(9,0.95)
(10,0.94)
(11,0.93)
(12,0.92)
(13,0.92)
(14,0.91)
(15,0.91)
(16,0.9)
(17,0.9)
(18,0.9)
(19,0.9)
(20,0.9)
(21,0.89)
(22,0.89)
(23,0.89)
(24,0.89)
(25,0.89)
(26,0.89)
(27,0.89)
(28,0.89)
(29,0.89)
(30,0.89)
(31,0.89)
(32,0.89)
(33,0.89)
(34,0.89)
(35,0.89)
(36,0.89)
(37,0.89)
(38,0.89)
(39,0.89)
(40,0.89)
(41,0.89)
(42,0.89)
(43,0.89)
(44,0.89)
(45,0.89)
(46,0.89)
(47,0.89)
(48,0.89)
(49,0.89)
(50,0.89)
(51,0.89)
(52,0.89)
(53,0.89)
(54,0.89)
(55,0.89)
(56,0.89)
(57,0.89)
(58,0.89)
(59,0.89)
(60,0.89)
(61,0.89)
(62,0.89)
(63,0.89)
(64,0.89)
(65,0.89)
(66,0.89)
(67,0.89)
(68,0.89)
(69,0.89)
(70,0.89)
(71,0.89)
(72,0.89)
(73,0.89)
(74,0.89)
(75,0.89)
(76,0.89)
(77,0.89)
(78,0.89)
(79,0.89)
(80,0.89)
(81,0.89)
(82,0.89)
(83,0.89)
(84,0.89)
(85,0.89)
(86,0.89)
(87,0.89)
(88,0.89)
(89,0.89)
(90,0.89)
(91,0.89)
(92,0.89)
(93,0.89)
(94,0.89)
(95,0.89)
(96,0.89)
(97,0.89)
(98,0.89)
(99,0.89)
};

\addplot[color=violet,line width = 1pt] plot coordinates { 
(1,1)
(2,1)
(3,1)
(4,1)
(5,1)
(6,0.99)
(7,0.98)
(8,0.97)
(9,0.96)
(10,0.96)
(11,0.95)
(12,0.95)
(13,0.95)
(14,0.95)
(15,0.94)
(16,0.94)
(17,0.94)
(18,0.94)
(19,0.94)
(20,0.94)
(21,0.94)
(22,0.94)
(23,0.94)
(24,0.94)
(25,0.94)
(26,0.94)
(27,0.94)
(28,0.94)
(29,0.94)
(30,0.94)
(31,0.94)
(32,0.94)
(33,0.93)
(34,0.93)
(35,0.93)
(36,0.93)
(37,0.93)
(38,0.93)
(39,0.93)
(40,0.93)
(41,0.93)
(42,0.93)
(43,0.93)
(44,0.93)
(45,0.93)
(46,0.93)
(47,0.93)
(48,0.93)
(49,0.93)
(50,0.93)
(51,0.93)
(52,0.93)
(53,0.93)
(54,0.93)
(55,0.93)
(56,0.93)
(57,0.93)
(58,0.93)
(59,0.93)
(60,0.93)
(61,0.93)
(62,0.93)
(63,0.93)
(64,0.93)
(65,0.93)
(66,0.93)
(67,0.93)
(68,0.93)
(69,0.93)
(70,0.93)
(71,0.93)
(72,0.93)
(73,0.93)
(74,0.93)
(75,0.93)
(76,0.93)
(77,0.93)
(78,0.93)
(79,0.93)
(80,0.93)
(81,0.93)
(82,0.93)
(83,0.93)
(84,0.93)
(85,0.93)
(86,0.93)
(87,0.93)
(88,0.93)
(89,0.93)
(90,0.93)
(91,0.93)
(92,0.93)
(93,0.93)
(94,0.93)
(95,0.93)
(96,0.93)
(97,0.93)
(98,0.93)
(99,0.93)
};
\addplot[color=blue,line width = 1pt] plot coordinates { 
(1,1)
(2,1)
(3,1)
(4,1)
(5,1)
(6,0.98)
(7,0.96)
(8,0.94)
(9,0.93)
(10,0.92)
(11,0.91)
(12,0.9)
(13,0.9)
(14,0.89)
(15,0.89)
(16,0.89)
(17,0.88)
(18,0.88)
(19,0.87)
(20,0.87)
(21,0.87)
(22,0.87)
(23,0.86)
(24,0.86)
(25,0.86)
(26,0.86)
(27,0.86)
(28,0.86)
(29,0.85)
(30,0.85)
(31,0.85)
(32,0.85)
(33,0.85)
(34,0.85)
(35,0.85)
(36,0.85)
(37,0.85)
(38,0.85)
(39,0.84)
(40,0.85)
(41,0.84)
(42,0.84)
(43,0.84)
(44,0.84)
(45,0.84)
(46,0.84)
(47,0.84)
(48,0.84)
(49,0.84)
(50,0.84)
(51,0.84)
(52,0.84)
(53,0.84)
(54,0.84)
(55,0.84)
(56,0.84)
(57,0.84)
(58,0.84)
(59,0.84)
(60,0.84)
(61,0.84)
(62,0.83)
(63,0.83)
(64,0.83)
(65,0.83)
(66,0.83)
(67,0.83)
(68,0.83)
(69,0.83)
(70,0.83)
(71,0.83)
(72,0.83)
(73,0.83)
(74,0.83)
(75,0.83)
(76,0.83)
(77,0.83)
(78,0.83)
(79,0.83)
(80,0.83)
(81,0.83)
(82,0.83)
(83,0.83)
(84,0.83)
(85,0.83)
(86,0.83)
(87,0.83)
(88,0.83)
(89,0.83)
(90,0.83)
(91,0.83)
(92,0.83)
(93,0.83)
(94,0.83)
(95,0.83)
(96,0.83)
(97,0.83)
(98,0.83)
(99,0.83)
};

  \end{axis}
 \end{tikzpicture}}

  \caption{\label{fig:energyconsumption}Accumulated energy consumption of both proposed predictors, normalized to the consumption of a naive, reactive \ac{DPM} $E^\text{DPM}$ (see ~\cref{energyestimation}), for the simulated traces.
    Similar to~\cref{fig:fnrreinffig}, reinforcement results are depicted as smooth, supervised results as dashed lines. 
    The scenarios are: $\vec{l}_\text{min}$ (blue), $\vec{l}_\text{avg}$ (green), and $\vec{l}_\text{max}$ (violet). 
}

\end{figure}
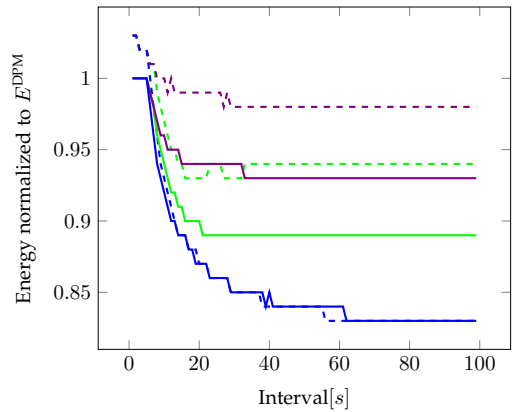

Finally, we evaluate the effectiveness of both approaches regarding the stated goal of achieving a reduction of the overall modem energy consumption.
\cref{fig:energyconsumption} shows the energy consumption for the (i) reinforcement and the (ii) supervised predictor, normalized to the (iii) state-of-the-art reactive \ac{DPM} for the same traces as ~\cref{fig:fnrreinffig}.

The most apparent difference, as explained in~\cref{subsec:metrics}, is the impact of $D_\text{DLG}$ on savable energy.
Secondly, we observe the added energy consumption of $E_Q$ from~\cref{ecogformula}, that leads to higher energy consumption both during the learning phase and during traffic intense intervals, where the modem cannot be turned off without missing a grant.
In the worst possible case of a failure to learn grant patterns this overhead would not be mitigated.
As shown in ~\cref{sec::predAccuracy}, in the simulations we performed, this did not occur.
After finishing the learning phase, both approaches quickly compensate for this energy overhead.
Indeed, both predictors for all scenarios enter the exploitation phase and achieve energy savings compared to the naive approach within $6$ and $11$~s.
In terms of achievable energy savings, the reinforcement predictor is generally superior.
Only for $\vec{l}_\text{min}$ the supervised predictor achieves the same energy saving of 17\%.
This is significant due to the higher inherent energy consumption $E_Q$, indicating a generally better performance of exploited opportunities for energy saving for this trace.
For the other two traces, the reinforcement predictor achieves 11\% ($\vec{l}_\text{avg}$) and 7\% ($\vec{l}_\text{max}$).
Compared to this, the supervised predictor exhibits ceilings of 6\% ($\vec{l}_\text{avg}$) and 2\% ($\vec{l}_\text{max}$).

Finally, one can observe the impact of trace characteristics, like downlink grant densities, for static traces.
In the long run, the trace that is best in terms of achievable \ac{FNR} ($\vec{l}_\text{max}$), with $D_\mathrm{DLG} = 0.82$, turns out to be the worst in terms of achievable energy savings.

\section{Conclusion}\label{conclusion}

This paper presents an approach for predictive dynamic power management for mobile devices in \ac{LTE} through grant prediction.
Two different approaches based on supervised learning and reinforcement learning that are optimized for accurate offline and efficient online learning, respectively, are investigated.
Moreover, we use a consistent complexity analysis to derive a comparable power model for both approaches.
Thus, using an identical environmental stimulus derived from a relevant simulated application model we perform a fair evaluation of both approaches.

As a result, both approaches need an interval of stable trace behavior to learn the grant patterns.
We believe the supervised predictor is the preferable solution if either the traffic density is low enough, or if more training of false negatives can be performed at early, offline design stages, as lower \acp{FNR} are achievable.
If neither trace stability can be guaranteed, nor offline learning improved, we argue for the reinforcement predictor due to its lower energy overhead and the observed online learning capabilities.
For longer stable scenarios, both approaches may achieve up to 17\% energy savings, with the reinforcement predictor providing generally higher savings.

Common to both approaches, the prediction-based DPM only gets activated if/once an acceptable error rate is achieved in the learning phase.
In general, the supervised predictor exhibits a lower margin for prediction errors, if the training can be performed on representative data, leading to less missed transmissions, at the cost of a higher inherent energy consumption.
In contrast to this, the reinforcement predictor achieves higher energy savings, at the cost of a slower learning speed.
For real-world applications, one has to find a balance between energy savings (preferring the reinforcement approach) and the least impact on service quality (in favor of supervised learning).
This might best be achieved through a combination of both approaches, where the supervised predictor is trained extensively offline, while the reinforcement predictor is employed for unexperienced trace scenarios.

\ifCLASSOPTIONcaptionsoff
  \newpage
\fi

\bibliographystyle{IEEEtran}

\bibliography{literature}

\end{document}